\documentclass[prd,english,floatfix,reprint,twocolumn,showpacs,superscriptaddress,longbibliography,12pt,tightenlines]{revtex4-1}
\usepackage{babel}
\usepackage[utf8]{inputenc}
\usepackage{geometry}
\usepackage[dvipsnames]{xcolor}
\usepackage{tabularx} 
\usepackage{amsmath}  
\usepackage{amssymb}
\usepackage{braket}
\usepackage{physics}
\usepackage{dsfont}
\usepackage{graphicx, wrapfig} 
\usepackage{multirow}
\usepackage[final]{hyperref} 
\hypersetup{
	colorlinks=true,       
	linkcolor=red,        
	citecolor=red,        
	filecolor=magenta,     
	urlcolor=blue         
}
\usepackage{csquotes}
\usepackage{blindtext}
\usepackage{float}
\usepackage{bm}
\usepackage{bigints}
\usepackage{empheq}
\usepackage{subcaption}

\numberwithin{equation}{section}
\numberwithin{figure}{section}
\numberwithin{table}{section}

\begin{document}

\title{Collapse instability and staccato decay of oscillons in various dimensions}
\author{B. C. Nagy}
\email{nagyboti98@gmail.com}
\affiliation{BME ``Momentum'' Statistical Field Theory Research Group}
\affiliation{Department of Theoretical Physics, Budapest University of Technology and Economics}
\author{G. Tak\'acs}
\email{takacs.gabor@ttk.bme.hu}
\affiliation{MTA-BME Quantum Correlations Group (ELKH)
\\Budafoki {\'u}t 6-8., 1111 Budapest, Hungary}
\affiliation{Department of Theoretical Physics, Budapest University of Technology and Economics}
\date{29th September 2021}

\begin{abstract}
    Oscillons are long-lived, slowly radiating solutions of nonlinear classical relativistic field theories. Recently it was discovered that in one spatial dimension their decay may proceed in "staccato" bursts. Here we perform a  systematic numerical study to demonstrate that although this behaviour is not confined to one spatial dimension, it quickly becomes unobservable when the dimension of space is increased, at least for the class of potentials considered here. To complete the picture we also present explicit results on the dimension dependence of the collapse instability observed for three-dimensional oscillons.
\end{abstract}
\maketitle
%
%

\section{Introduction}

Nondissipative configurations play a very important role in classical field theory. These are solutions of the equation of motion, for which the energy density remains localised during the time evolution. For a wide class of nondissipative configurations, their stability is guaranteed by some conserved charge. Such charges may be topological, or also higher-spin charges due to integrability. However, there are also examples of nontopological stable soliton configurations such as the so-called "Q-balls" in complex scalar fields whose stability is guaranteed by the global $U(1)$ charge \cite{COLEMAN1985263}, and also "I-balls" whose stability is guaranteed by adiabatic invariance  \cite{KASUYA200399}.

Surprisingly however, even in theories where such conserved charges or adiabatic mechanisms don't exist, it is still possible to find metastable solutions, which are strictly speaking dissipative, but the energy dissipates very slowly compared to the characteristic dynamical time scales. These are spatially localised, coherently oscillating, long-living solutions of relativistic classical scalar field theories, known as oscillons. What makes them even more interesting is that they are not exceptional configurations, but attractors in the space of field configurations \cite{PhysRevD.52.1920,Saffin_2007,Zhang_2020,PhysRevLett.101.011602,PhysRevD.80.125037,Cyncynates_2021}; moreover, they are stable against small perturbations \cite{PhysRevD.49.2978}. The main conditions for their existence are that the initial configuration's energy is higher than a threshold value (which is characteristic to the given model), and that the model is nonlinear (for an estimation of the lifetime of localised lumps in linear theories see \cite{PhysRevD.52.1920}). Since their occurrence is more or less independent of the details of the theory, they are widespread in models of cosmology and particle physics. Their properties have also been studied at the quantum level \cite{Hertzberg_2010,Saffin_2007}. For a comprehensive recent review of oscillons the interested reader is referred to \cite{fodor2019review}.

The dimensionality of space-time is known to strongly influence the dynamics of oscillons. For a theory with a scalar particle of mass $m$, their periods are $\mathcal{O}(10)\cdot m^{-1}$, while their typical lifetimes in three spatial dimensions are usually $\mathcal{O}(10^3) \cdot m^{-1}$ \cite{PhysRevD.52.1920}, and after slowly radiating their energy away, their final decay happens via a sudden collapse. Oscillons oscillate with a frequency below the threshold $m$, which increases as they radiate their energy away, and the sudden collapse happens when their frequency approaches the threshold. However, in one and two spatial dimensions, oscillons are radiating even slower: in two dimensions, their typical lifetime exceeds $\mathcal{O}(10^7)\cdot m^{-1}$ \cite{PhysRevD.74.105005,PhysRevE.62.1368}, and no collapse of oscillons has been observed. Earlier studies mainly focused on the time evolution of oscillons with frequencies close to the mass threshold, whose dynamics can be described analytically using the  small amplitude expansion \cite{PhysRevD.78.025003,fodor2019review}; it is also possible to find nonperturbative radiative corrections \cite{Fodor_2009_1,Fodor_2009_2}. 

More recently it was found that in a particular class of field theories in one spatial dimension large amplitude oscillons decay in bursts via the so-called "staccato" mechanism \cite{Dorey_2020}. These bursts occur when the monotonically increasing frequency of the oscillon reaches a value for which a higher harmonic crosses the threshold and therefore changes its nature from localised to radiative mode. Given the apparent generality of the above condition, it is natural to wonder whether this staccato decay can be found in higher spatial dimensions as well.

In the present work we study the decay mechanisms of oscillons when the number of spatial dimensions is increased above one. In Sec. \ref{sec:prelim} we introduce the essential concepts and explain the choice of the initial configuration (the core of a so-called quasibreather configuration \cite{PhysRevD.74.124003}) for the time evolution. In Sec. \ref{sec:nummethods} we present and discuss the numerical methods used to construct the initial configuration and to compute the time evolution. In Sec. \ref{sec:results} we present the results of our numerical investigations, starting with a validation of our methods by reproducing dynamics in three spatial dimensions and the proceeding to the case of interest. Finally we present our conclusions in Sec. \ref{sec:conclusions}.

\section{Preliminaries}\label{sec:prelim}

We consider relativistic classical field theories of a single scalar field in $D$ spatial dimensions, defined by the action
\begin{equation}
    S = \int \mathrm{d}^{D+1}x\ \mathcal{L}\ ,
    \label{action}
\end{equation}
with the Lagrangian density
\begin{equation}
    \mathcal{L} = \frac{1}{2} \left(\frac{\partial \phi}{\partial t}\right)^2 - \frac{1}{2} \left( \bm{\nabla}\phi\right)^2 - V(\phi)\ .
\end{equation}
The Euler-Lagrange equation of motion for the field $\phi$ is given by
\begin{equation}
    \frac{\partial^2 \phi}{\partial t^2} - \Delta \phi = -V'(\phi)\ .
    \label{Euler_Lagrange}
\end{equation}
We assume that the potential has a minimum at $\phi=0$ and denote the mass of elementary excitations by $m^2=V''(0)$. Oscillons are spatially localised, almost exactly time-periodic solutions admitted by Eq. (\ref{Euler_Lagrange}),  which can only exist when the frequency $\omega$ of their quasiperiodic motion satisfies $\omega<m$.

In the following we restrict ourselves to spherically symmetric field configurations, for which the equation of motion can be simplified as
\begin{equation}
    \frac{\partial^2 \phi}{\partial t^2} - \frac{\partial^2 \phi}{\partial r^2} - \frac{D-1}{r} \frac{\partial \phi}{\partial r} = -V'(\phi)\ .
    \label{EOM}
\end{equation}
Note that spherical symmetry allows treating the spatial dimension $D$ as a continuous parameter, which we use to our advantage below.

The canonical energy density of a spherically symmetric field configuration is given by
\begin{equation}
    \mathcal{E} = \frac{1}{2}\left( \frac{\partial\phi}{\partial t}\right)^2 + \frac{1}{2}\left( \frac{\partial\phi}{\partial r}\right)^2 + V(\phi)\ ,
    \label{energy_density}
\end{equation}
and the energy contained in a sphere of radius $R$ can be computed as
\begin{equation}
    E(R) = \frac{2\pi^{{D}/{2}}}{\Gamma\left({D}/{2}\right)} \int_0^R \mathrm{d}r\ r^{D-1}\ \mathcal{E}\ .
    \label{energy}
\end{equation}
It is also possible to define the effective radius of a field configuration
\begin{equation}
    R_{eff} = \frac{\int_0^R \mathrm{d}r\ r^{D}\ \mathcal{E}}{\int_0^R \mathrm{d}r\ r^{D-1}\ \mathcal{E}}\ .
    \label{effrad}
\end{equation}
which characterises the spatial localisation of the energy density.

In general, oscillons are not exactly time periodic, but lose energy via radiation into modes of the scalar field which can be considered free waves at spatial infinity; this results in the increase of the frequency of the oscillon core. A closely related concept to oscillons is that of quasibreathers \cite{PhysRevD.74.124003}, which are exactly periodic configurations with a standing wave tail of radiation which compensates for the energy loss. The price of stabilising the solutions is that the standing wave tail, albeit of small amplitude, nevertheless results in infinite total energy; therefore a quasibreather configuration is, strictly speaking, nonphysical. However, they are extremely useful objects to study, as the time evolution of the oscillon core can be well represented as adiabatic time evolution through different frequency quasibreathers.

It is possible to find oscillon configurations starting a time evolution from essentially any well-localised scalar field configuration, e.g., a simple Gaussian profile. However, in general these configurations quickly radiate a part of their energy away before settling down into the eventual oscillon configuration, which also means that their initial amplitude and frequency are generally not under control. To avoid these issues, it is much better to construct a quasibreather solution first, and start with its core obtained by discarding the standing wave tail. This leads to a substantially cleaner evolution as well as enabling a much better control of the initial amplitude and frequency. There exists a number of numerical methods for the construction of quasibreathers \cite{Alfimov_2000,Saffin_2007,PhysRevD.74.124003,Zhang_2020}, all of which have their pros and cons; here we use a systematical approach which allows finding quasibreathers in a wide range of frequencies below the mass threshold for any spatial dimensions $D$.

\section{Numerical methods}\label{sec:nummethods}

In this ection we present our numerical methods for finding the quasibreather profile for a given frequency $\omega$ in dimension $D$, and then time evolving this quasibreather initial configuration. The concrete numerical computations are carried out in a deformed sine-Gordon model (motivated by \cite{Dorey_2020}), where the potential is \footnote{We set the particle mass $m=1$ for any $\lambda$, giving us a natural set of units for energy, time and distance.}
\begin{equation}
    V(\phi,\lambda)=(1-\lambda)(1-\cos\phi) + \frac{\lambda}{8\pi^2}\phi^2(\phi-2\pi)^2\ ,
\end{equation}
which in $D=1$ and for $\lambda=0$ permits an exactly periodic solution for $\omega \in (0,1)$, the sine-Gordon breather
\begin{equation}
    \phi^{sG}(t,x) = 4\arctan(\frac{\epsilon \cos\omega t}{\omega \cosh \epsilon x})\ ,
\label{sgbreather}\end{equation}
where the parameter $\epsilon=\sqrt{1-\omega^2}$ is essentially the amplitude of the breather.

We note here, that the fact that we assume spherical symmetry implies that $D$ is just a parameter in Eq. (\ref{EOM}), and therefore can be any real number, and doesn't necessarily need to be an integer. 

\subsection{Constructing the quasibreather profile}\label{subsec:qbprofile}
\begin{figure*}
\centering
\begin{subfigure}{.5\textwidth}
  \centering
  \includegraphics[width=1\linewidth]{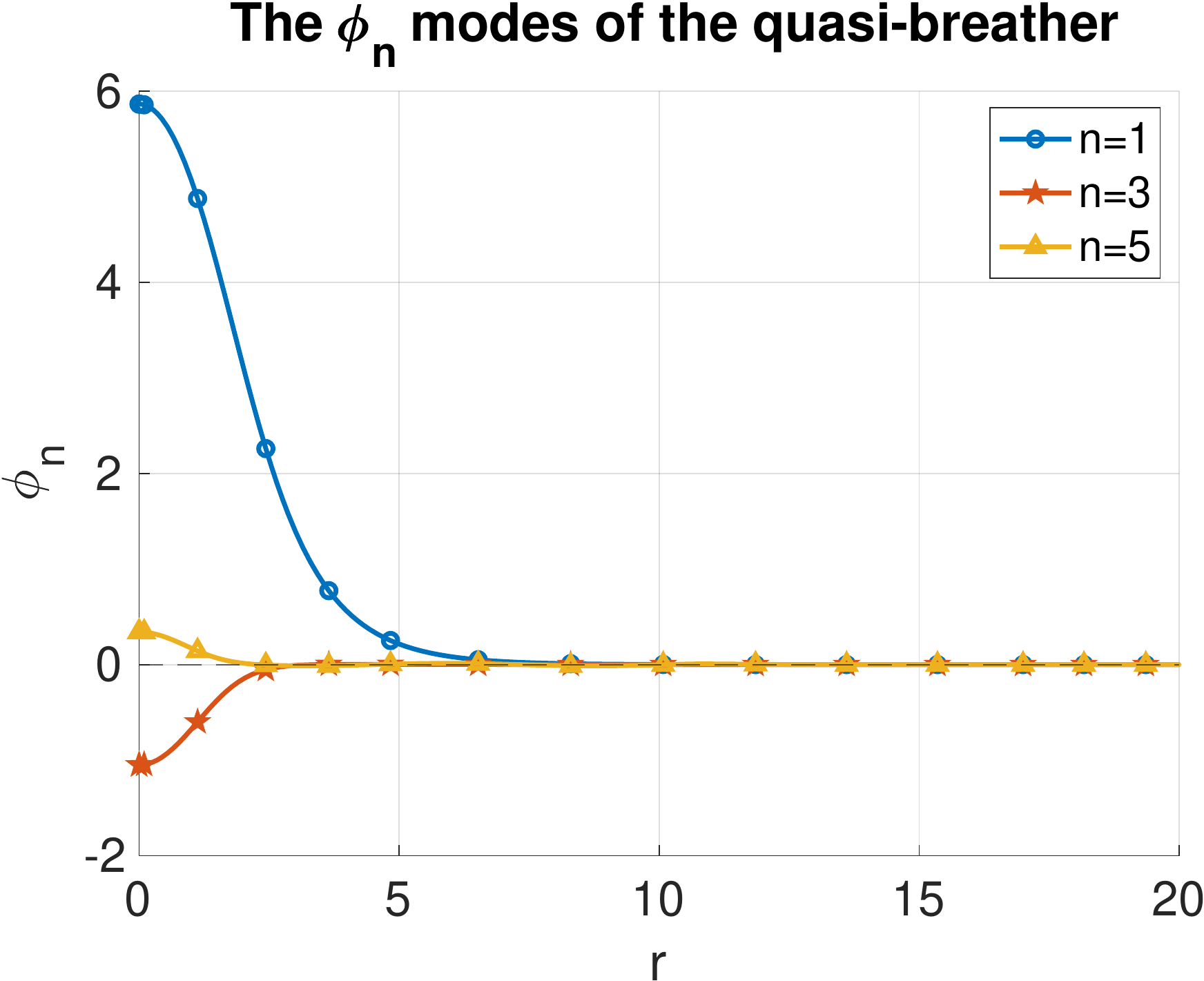}
  \caption{\label{fig:sub1}}
\end{subfigure}%
\begin{subfigure}{.5\textwidth}
  \centering
  \includegraphics[width=1\linewidth]{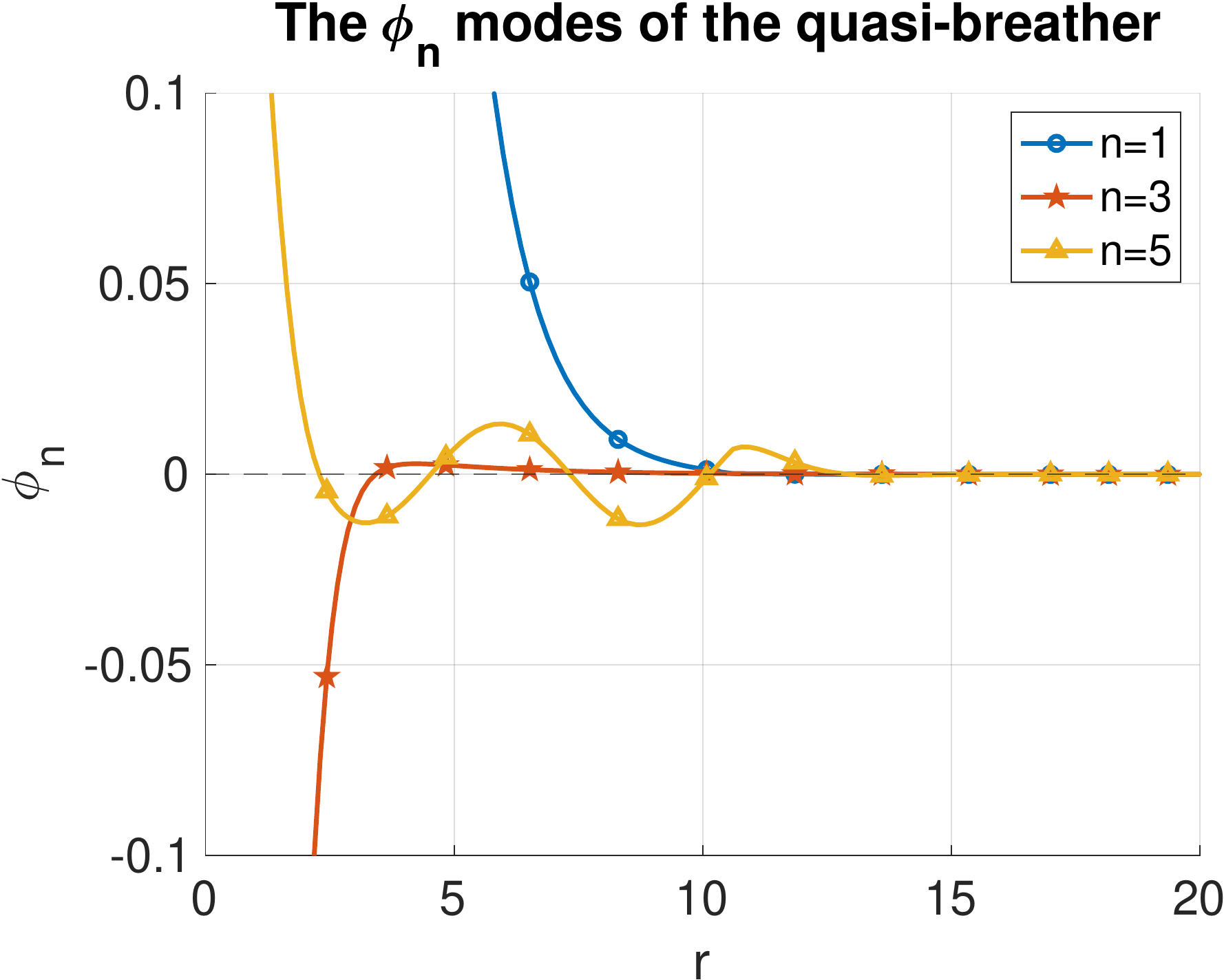}
  \caption{\label{fig:sub2}}
\end{subfigure}
\caption{\label{fig:test}The quasibreather for $\omega=0.3$, $D=1$, $\lambda=0$, for $N=5$, including the exponential suppression factor \eqref{exp_supression} with $r_0=10/\sqrt{1-\omega^2}$ (note that even modes are missing due to parity).}
\end{figure*}

Our method is based on \cite{Alfimov_2000} with minor modifications; here we only describe the essential points. A quasibreather is a spatially localised, exactly time-periodic solution of Eq. (\ref{EOM}) which oscillates coherently around the $\phi=0$ vacuum; therefore we look for a solution of the form
\begin{equation}
    \phi(t,r) = \sum_{m=1}^N \phi_m(r)\cos(m\omega t)\ ,
    \label{Ansatz}
\end{equation}
where the expansion into higher harmonics is truncated at level $N$. Substituting the above ansatz into Eq. (\ref{EOM}) gives
\begin{eqnarray}
    &&\sum_{m=1}^N \left( m^2\omega^2 +\frac{\partial^2}{\partial r^2} + \frac{D-1}{r}\frac{\partial}{\partial r} \right)\phi_m(r) \cos(m\omega t) \nonumber\\
    &&= V'\left( \sum_{m=1}^N \phi_m(r)\cos(m\omega t) \right)\ ,
\end{eqnarray}
which after Fourier transformation can be cast into the form
\begin{eqnarray}
    &&\left( n^2\omega^2 +\frac{\partial^2}{\partial r^2} + \frac{D-1}{r}\frac{\partial}{\partial r} \right)\phi_n(r) 
    \\
    &&= \frac{\omega}{\pi} \int_0^{2\pi/\omega} V'\left( \sum_{m=1}^N \phi_m(r)\cos(m\omega t) \right) \cos(n\omega t)\ ,\nonumber
\end{eqnarray}
for $n=1,2,...,N$. This is a system of coupled differential equations which can be solved numerically upon specifying the boundary conditions. We require the solution to be nonsingular in the origin, and to be spatially localised, so we set the following boundary conditions,
\begin{equation}
    \frac{\mathrm{d}}{\mathrm{d}r}\phi_n(r)\biggr\rvert_{r=0} = 0\quad ,\quad
    \lim_{r\to \infty} \phi_n(r) = 0\ .
    \label{BC}
\end{equation}
In the asymptotic region where $\phi$ is small, the modes of Eq. (\ref{Ansatz}) decouple, and the large distance asymptotics of the modes of a field obeying Eq. (\ref{EOM}) are
\begin{eqnarray}
    \phi_n(r) &&\sim  {r^{\frac{1-D}{2}}}\exp(-\sqrt{1-n^2\omega^2}\cdot r)\nonumber\\ && \textrm{for } n\omega<1,\ \textrm{localised modes,} \nonumber\\
    \phi_n(r) &&\sim {r^{\frac{1-D}{2}}}\sin(\sqrt{n^2\omega^2-1}\cdot r+\varphi_n)
    \nonumber\\
    && \textrm{for } n\omega>1,\ \textrm{radiation modes.}
    \label{asymptotics}
\end{eqnarray}
The solution can be constructed by finding appropriate "initial conditions", $\phi_n(r=0)$, at the origin such that together with the other initial condition for the derivative, the asymptotic boundary condition at $r\rightarrow\infty$ is satisfied. Since the cases we consider have $\lambda\ll 1$, a good starting position for the $\phi_n(r=0)$ for $D=1$ is provided by the sine-Gordon breather Eq. (\ref{sgbreather}):
\begin{equation}
    \phi_n(0) \approx \phi^{sG}_n(0) = \frac{\omega}{\pi} \int_0^{2\pi/\omega} \mathrm{d}t \cos(n\omega t)\ \phi^{sG}(t,0)\ .
\end{equation}
The next step is finding a value for $\phi_1(0)$, for which $\phi_1$ -- which always decays in the considered frequency range, i.e., below the mass threshold -- obeys Eq. (\ref{asymptotics}), while keeping all $\phi_n(r=0)$ with $n>1$ fixed. It is argued in \cite{Alfimov_2000}, that such $\phi_1(0)$ is atypical in the sense, that $\phi_1$ usually converges to either a negative or a positive value in the limit $r\to \infty$; nevertheless, it is relatively easy to converge to a value $\phi_1(0)$ for which Eq. (\ref{asymptotics}) holds by the bisection method. The next step is to find a value for $\phi_2(0)$; however, this requires readjusting the value of $\phi_1(0)$ as well. It is possible to proceed by progressively including more and more modes, but in practice the procedure becomes exponentially slower with increasing number of modes, and so in our simulations we always stopped at two or three modes.

For the case $\lambda=0$, parity implies that even modes in Eq. (\ref{Ansatz}) vanish \cite{Zhang_2020,fodor2019review}, which results in a considerable gain in computing time. In addition, for the case $\lambda\ll 1$ it is often enough to compute only the odd modes (due to the even ones being negligible) because it still gives a suitably good approximation of the quasibreather solution. It is argued in \cite{Alfimov_2000}, that the oscillatory tail modes always oscillate around $\phi=0$ according to Eq. (\ref{asymptotics}), but in many cases the quasibreather solution can be improved further by adjusting $\phi_n(0)$ to minimise the amplitude to obtain a better quasibreather profile if necessary.

Once we have a quasibreather of frequency $\omega$ in dimension $D$, we can find a quasibreather of the same frequency in a slightly higher dimension, $D+\delta D$, starting from the initial guess
\begin{equation}
    \phi_n^{D+\delta D}(0) = \phi_n^D(0)\ ,
\end{equation}
and we adjust these values in the way described above.

One last subtlety is that the quasibreathers must be truncated at some spatial coordinate to obtain a localised, finite-energy initial configuration for numerical simulations of oscillon time evolution. For localised modes this is easy as they decay quickly enough; regarding the radiation modes, we enforce exponential decay in the asymptotic region (with a scale corresponding to the wave number of the given mode), i.e., we multiply the asymptotic behaviour in Eq. (\ref{asymptotics}) by 
\begin{equation}
    \exp(-\sqrt{n^2\omega^2-1}\cdot(r-r_0))
\label{exp_supression}\end{equation} for the radiation modes, where $r_0$ is an arbitrarily chosen truncation coordinate outside the core region. Our numerical studies show that the choice of $r_0$ for this exponential suppression only negligibly influences the temporal dynamics.

For the considerations of Sec. \ref{sec:results}, the crucial issue is to have control over the initial frequency of the evolving oscillon, and so the quality of the initial conditions obtained by the above procedure can be independently established by verifying that the time evolution of the oscillon starts suitably close to the frequency for which the quasibreather solution was constructed, providing an independent justification for our numerical procedure.

An example of a quasi-breather solution obtained by the above method is shown in Fig. \ref{fig:test}.

\subsection{Time evolution of oscillons}

Time evolution is computed using a method described in \cite{PhysRevD.74.124003, fodor2019review}, with some modifications. The original spatial coordinate $r\in[0,\infty)$ is mapped to the domain $R\in[0,1)$ by
\begin{equation}
    r = \frac{2R}{\kappa(1-R^2)}\ ,
\end{equation}
which helps in treating the boundary conditions and avoiding associated numerical instabilities. The parameter $\kappa$ controls the number of grid points in the core and in the radiation region; in our simulations we used $\kappa = 0.05$. Under this change of the variables Eq. (\ref{EOM}) becomes
\begin{eqnarray}
    \frac{\partial^2 \phi}{\partial t^2} =&& \frac{\kappa^2 (1-R^2)^3}{2(1+R^2)} \Big[ \frac{(1-R^2)}{2(1+R^2)}\frac{\partial^2 \phi}{\partial R^2} \label{numerical_equation}\\
    && - \frac{R(3+R^2)}{(1+R^2)^2} \frac{\partial \phi}{\partial R}
     + \frac{D-1}{2R} \frac{\partial \phi}{\partial R} \Big] - V'(\phi) \ .\nonumber
\end{eqnarray}
Introducing the new variables
\begin{equation}
    \phi_t = \frac{\partial \phi}{\partial t}\ , \hspace{1cm} \phi_R = \frac{\partial \phi}{\partial R}\ ,
\end{equation}
results in the following system of coupled differential equations
\begin{align}
    \frac{\partial \phi}{\partial t} =& \phi_t \ , \nonumber \\
    \frac{\partial \phi_t}{\partial t} =&  \frac{\kappa^2 (1-R^2)^3}{2(1+R^2)} \Big[ \frac{(1-R^2)}{2(1+R^2)}\frac{\partial \phi_R}{\partial R} 
    \nonumber \\
    &- \frac{R(3+R^2)}{(1+R^2)^2} \phi_R + \frac{d-1}{2R} \phi_R \Big] - V'(\phi) \ , \nonumber \\
    \frac{\partial \phi_R}{\partial t} =& \frac{\partial \phi_t}{\partial R} \ .\label{numeqtosolve}
\end{align}
Upon specifying the initial configuration $\phi(0,R)$ as the numerically obtained quasibreathers at time $t=0$, the system of Eqs. (\ref{numeqtosolve}) can be solved for $\phi(t,R)$ i.e., $\phi(t,r)$. For the temporal direction we chose a discretisation with $\Delta t = \Delta R$, and used a fourth-order Runge-Kutta method combined with the method of lines.

For numerical stability it is necessary to suppress the short-wavelength modes which can be achieved by including dissipative terms \cite{PhysRevD.77.025019} 
\begin{equation}
    \mathcal{D} = \mathcal{K} \left( \partial_R^6 \Phi\right) \left( \Delta R \right)^5\ ,
\end{equation}
in each equation in the system (\ref{numeqtosolve}), which only introduces numerical deviation at fifth order which is one higher than the order of the Runge-Kutta method used.

To satisfy the boundary conditions in Eq. (\ref{BC}) and to further increase stability, the system (\ref{numeqtosolve}) was solved in an extended region $R\in [-1-\epsilon,1+\epsilon]$, with $\phi$, $\phi_t$ and $\phi_R$ set to zero for  $|R| \geq 1$, and the fields were symmetrised after each time-step:
\begin{eqnarray}
    \phi(R) &&\rightarrow \frac{\phi(R)+\phi(-R)}{2}\ ,
    \nonumber \\ 
    \phi_t(R) &&\rightarrow \frac{\phi_t(R)+\phi_t(-R)}{2}\ ,
    \nonumber \\ 
    \phi_R(R) &&\rightarrow \frac{\phi_R(R) - \phi_R(-R)}{2}\ .
\end{eqnarray}
The energy and the effective radius can be calculated from Eqs. (\ref{energy},\ref{effrad}). The time sequence characterising the evolution of the frequency is determined from the time points $t_i^0$ when the field vanishes at the origin $\phi(R=0,t_i^0)=0$; the frequency $\omega_i$ at any given $t_i^0$ is defined from the time elapsed between the two adjoining time points
\begin{equation}
    \omega_i = \frac{2\pi}{t^0_{i+1}-t^0_{i-1}}\ .
    \label{freq}
\end{equation}

\section{Results}\label{sec:results}

In this section we present our results. First we consider the case of high frequency oscillons, where we investigated the collapse instabilities, and then the low frequency ones, where we studied the staccato decay mechanism.

\subsection{High-frequency regime}

\begin{figure}
    \centering
    \includegraphics[width=0.45\textwidth]{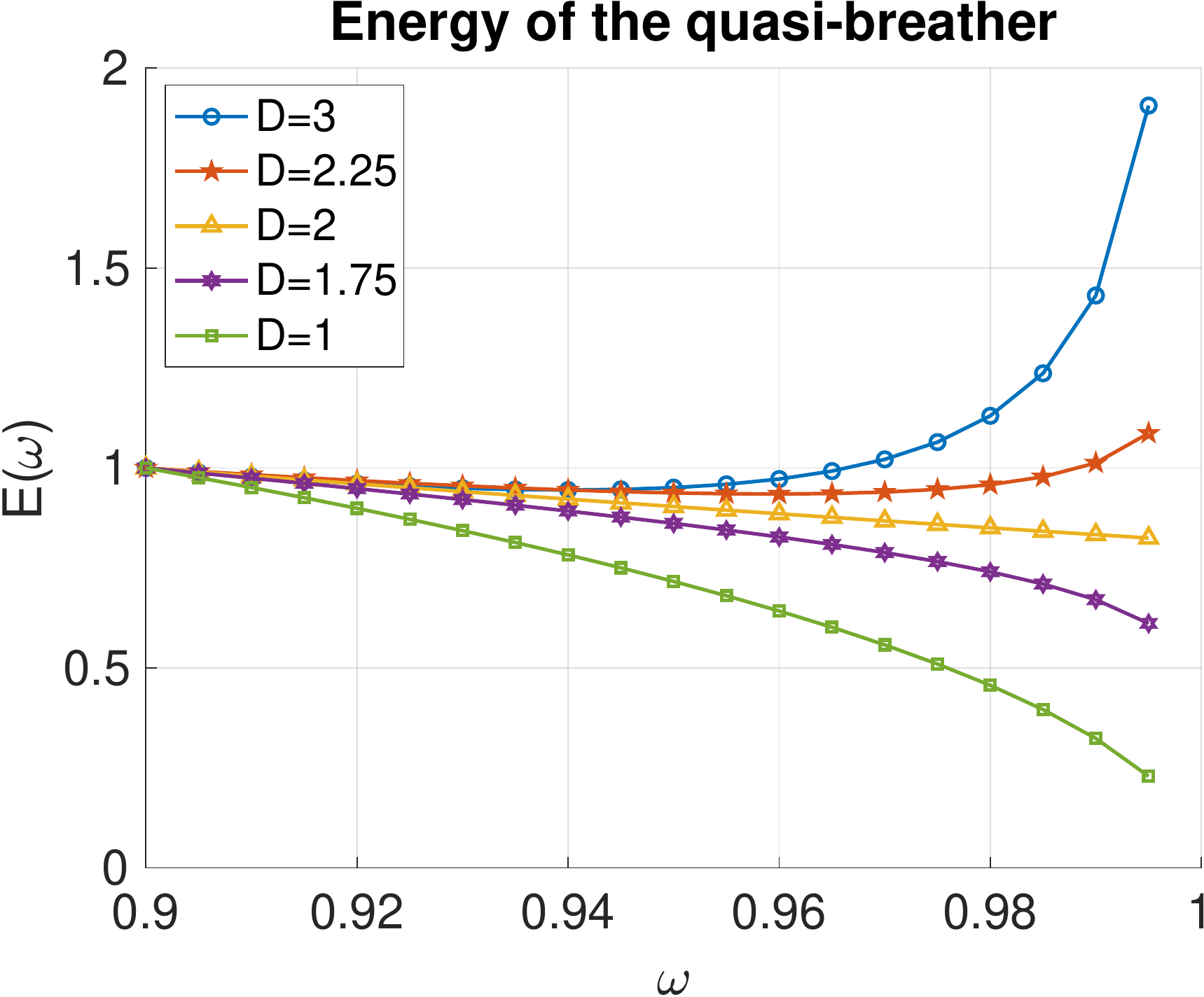}
    \caption{\label{energycurve} Quasibreather energy as a function of its frequency $\omega$ for different dimensions $D$. The curves obtained for $D>2$ have a well-defined minimum, while the ones corresponding to $D\leq 2$ are monotonically decreasing.}
\end{figure}

For this investigation we constructed quasibreather profiles for $\omega \in [0.9,1)$, and calculated their energy to obtain their $E(\omega)$ function for the sine-Gordon model ($\lambda=0$) in different dimensions, according to the definition of the energy described in Appendix \ref{QBenergy}. This energy curve has already been investigated previously \cite{PhysRevD.74.124003,Saffin_2007}, but systematic study of its dependence on $D$ has not been carried out yet. The energy-frequency curves for different values of the dimension $D$ were normalised to $E(\omega=0.9)=1$ to allow better comparison and are shown in Fig. \ref{energycurve}.

\begin{figure*}
\centering
\begin{subfigure}{.49\textwidth}
  \centering
  \includegraphics[width=1\linewidth]{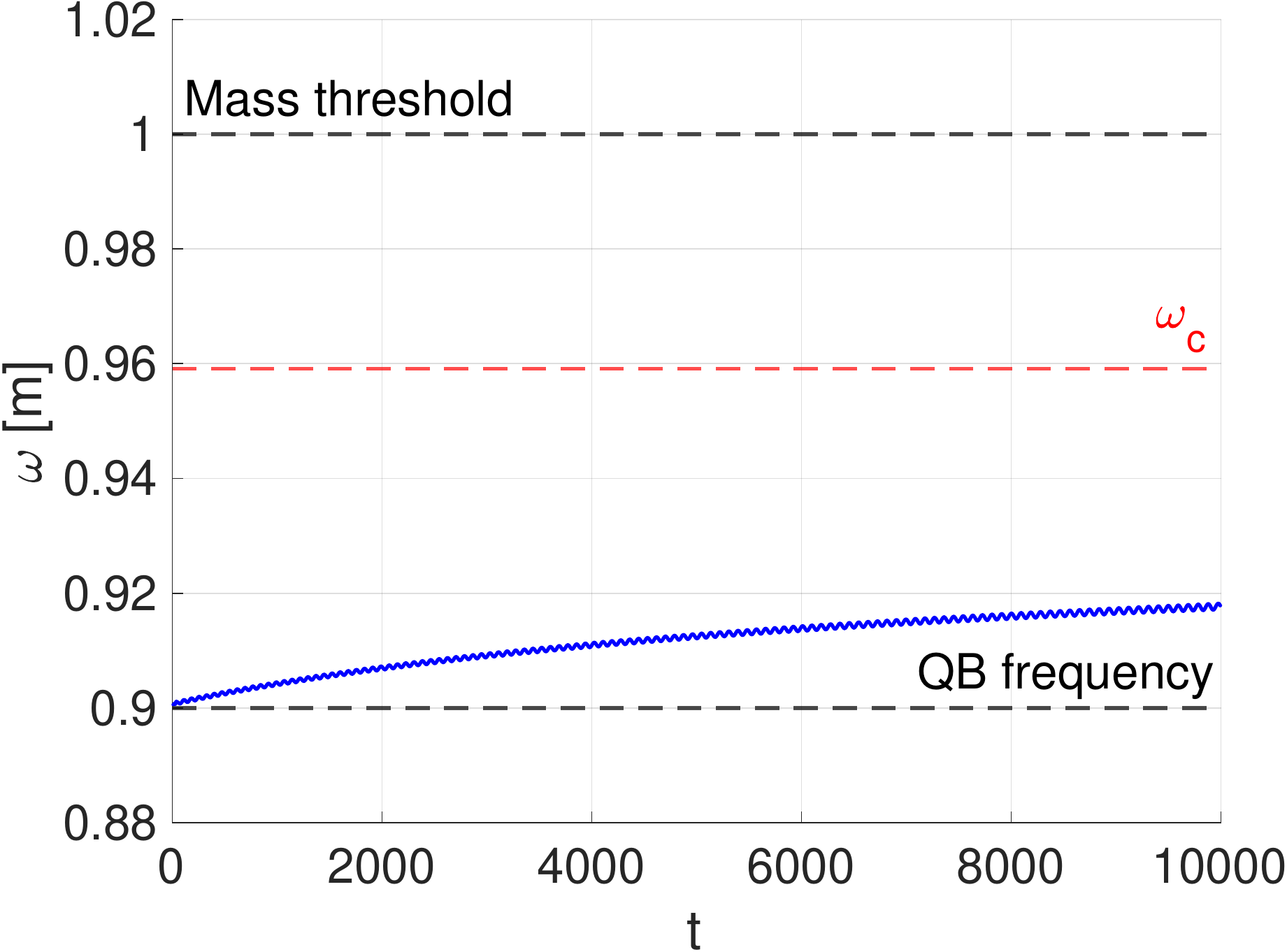}
  \caption{$D=2.25$}
\end{subfigure}%
\begin{subfigure}{.49\textwidth}
  \centering
  \includegraphics[width=1\linewidth]{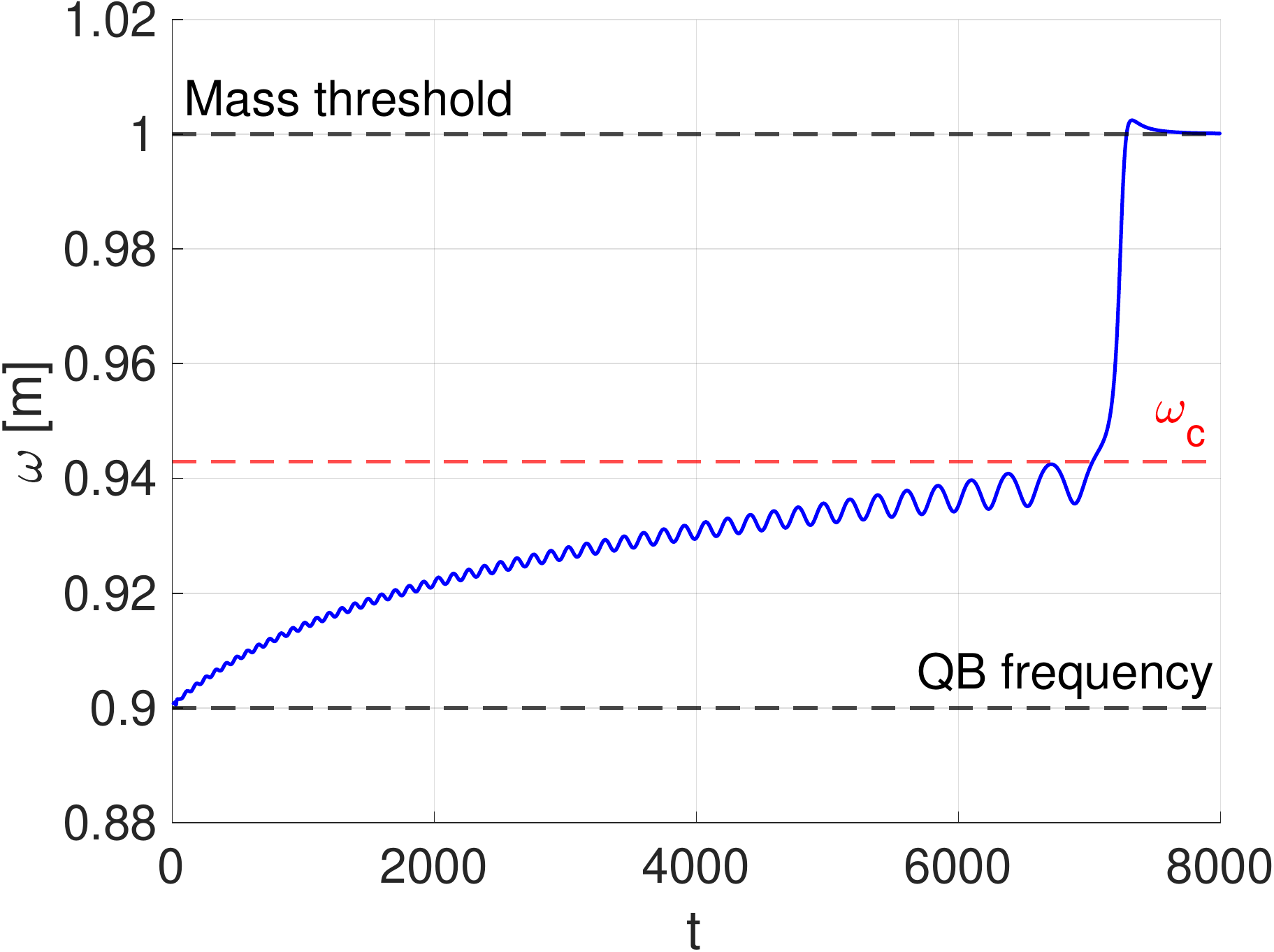}
  \caption{$D=2.50$}
\end{subfigure}%

\begin{subfigure}{.49\textwidth}
  \centering
  \includegraphics[width=1\linewidth]{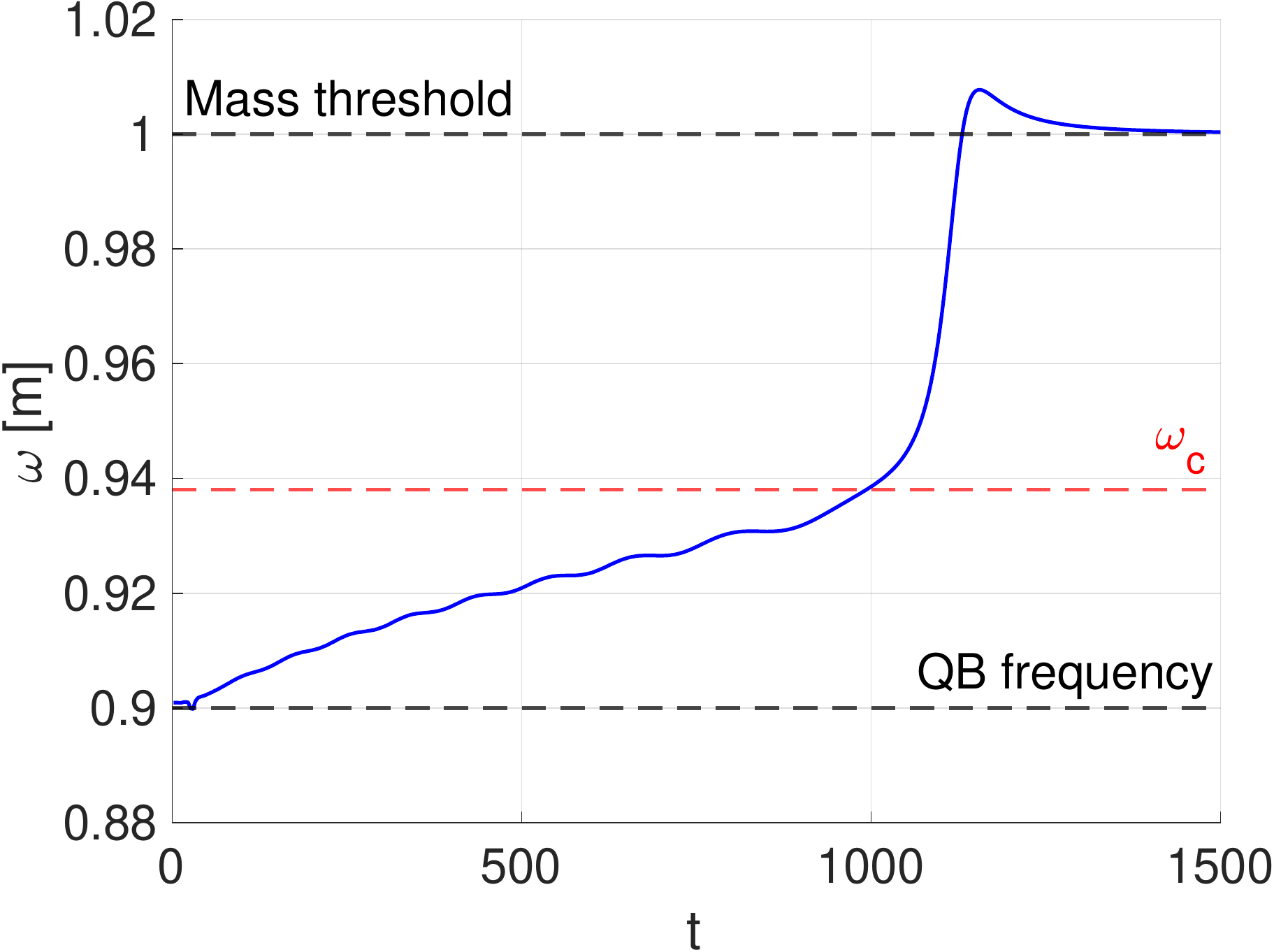}
  \caption{$D=2.75$}
\end{subfigure}
\begin{subfigure}{.49\textwidth}
  \centering
  \includegraphics[width=1\linewidth]{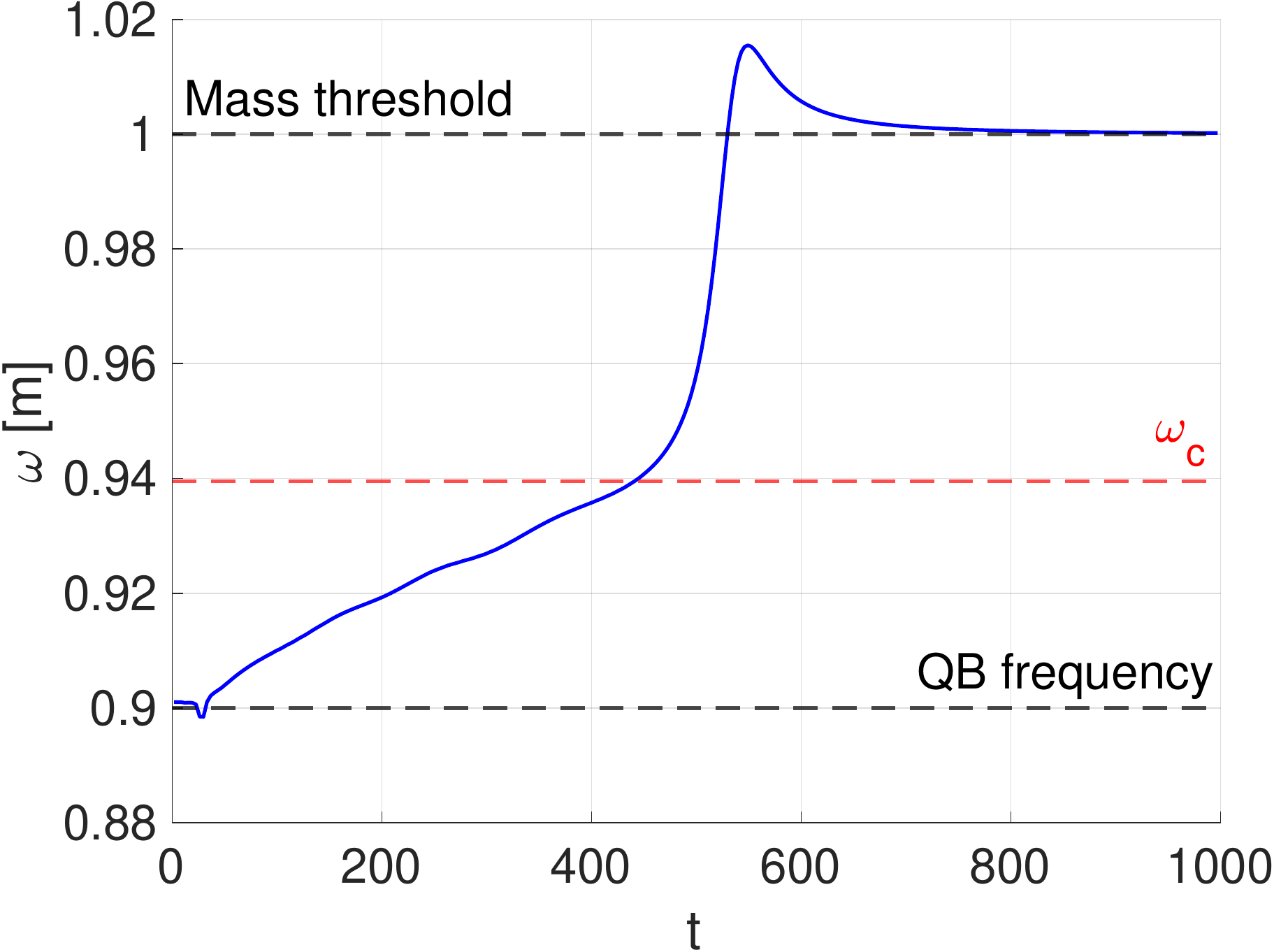}
  \caption{$D=3.00$}
\end{subfigure}
\caption{\label{frequencies} Time evolution of the frequency of quasibreathers from $\omega=0.9$, $N=3$ in different dimensions. Note the different time scales.}
\end{figure*}

The sudden collapse of oscillons for $D=3$ is due to the existence of a minimum of the $E(\omega)$ function at some frequency $\omega_c<1$. The radiating oscillon loses energy by a continuous radiation while the frequency is gradually increasing towards the threshold $\omega=1$. However, gradual emission of radiation cannot increase the frequency beyond $\omega_c$, while the dissolution of the oscillon eventually implies its decay to radiation modes with frequencies $\omega\geq 1$. As a result, the final decay of the oscillon eventually proceeds by a sudden collapse instead of gradual emission of radiation, which can be clearly identified in the frequency curves shown in Fig. \ref{frequencies}. The dependence of the critical frequency on $D$ is shown in Table \ref{critfreq}; the position of the minimum is obtained from a second-order polynomial fit around the minimal point of the curves for $D>2$.

\begin{table}
\centering
\begin{tabular}{c|cccc}
$D$      & $2.25$   & $2.50$   & $2.75$   & $3.00$   \\
\hline
$\omega_c$ & $0.9591$ & $0.9429$ & $0.9381$ & $0.9395$
\end{tabular}
\caption{\label{critfreq} Frequency of the minimum of the energy curve in different dimensions.}
\end{table}

It is also clear from Fig. \ref{energycurve} that for $D \leq 2$ the minimum ceases to exist, and the $E(\omega)$ curve becomes monotonically decreasing. In $D \leq 2$ the oscillon then continues to radiate away indefinitely without displaying any sudden collapse instability. This behaviour was already suggested in \cite{PhysRevD.78.025003} from the small amplitude expansion of the $\phi^4$ model, but here we numerically obtained the curves also in the sine-Gordon model.

\begin{figure*}
\centering
\begin{subfigure}{.99\textwidth}
  \centering
  \includegraphics[width=.49\linewidth]{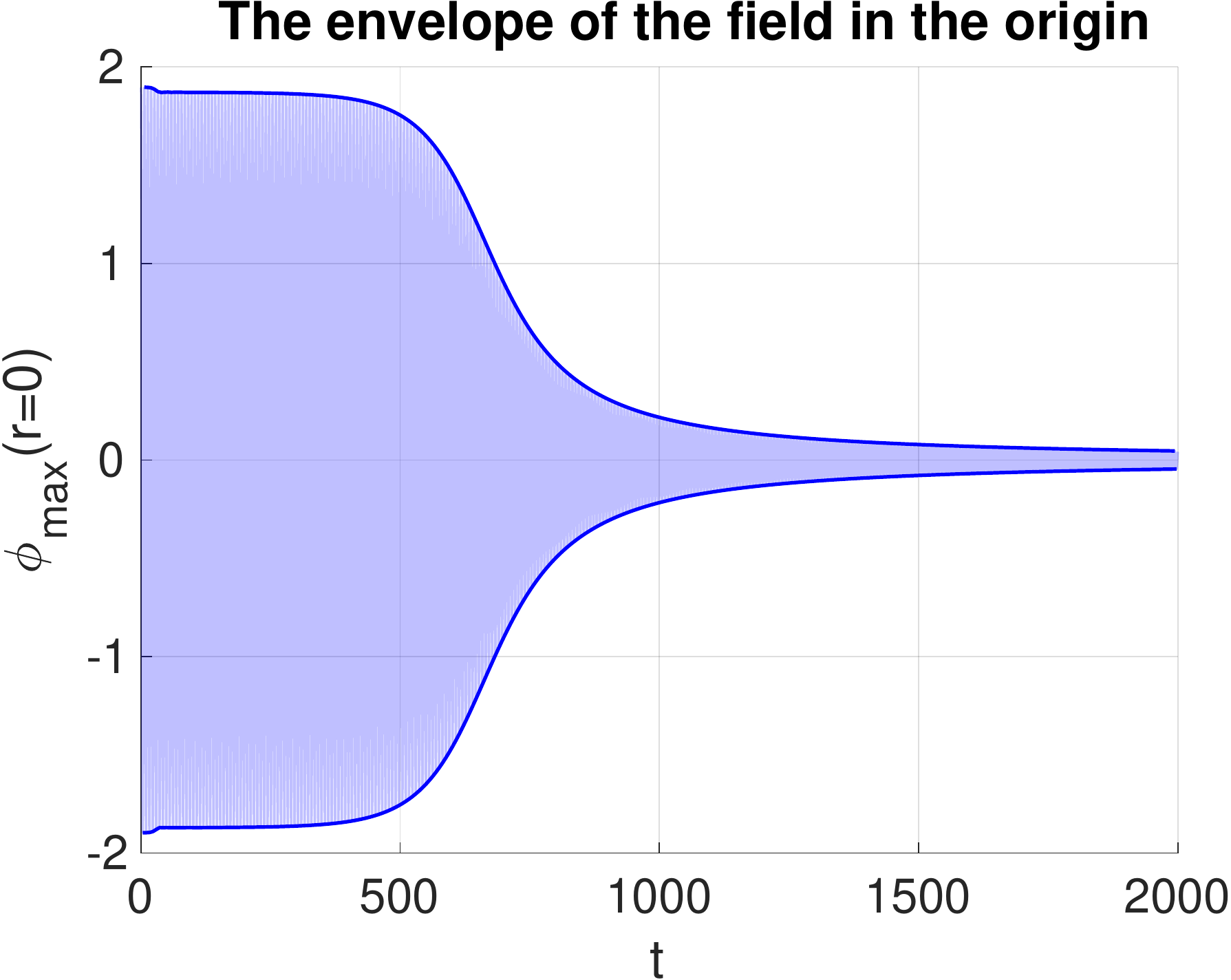}
  \includegraphics[width=.49\linewidth]{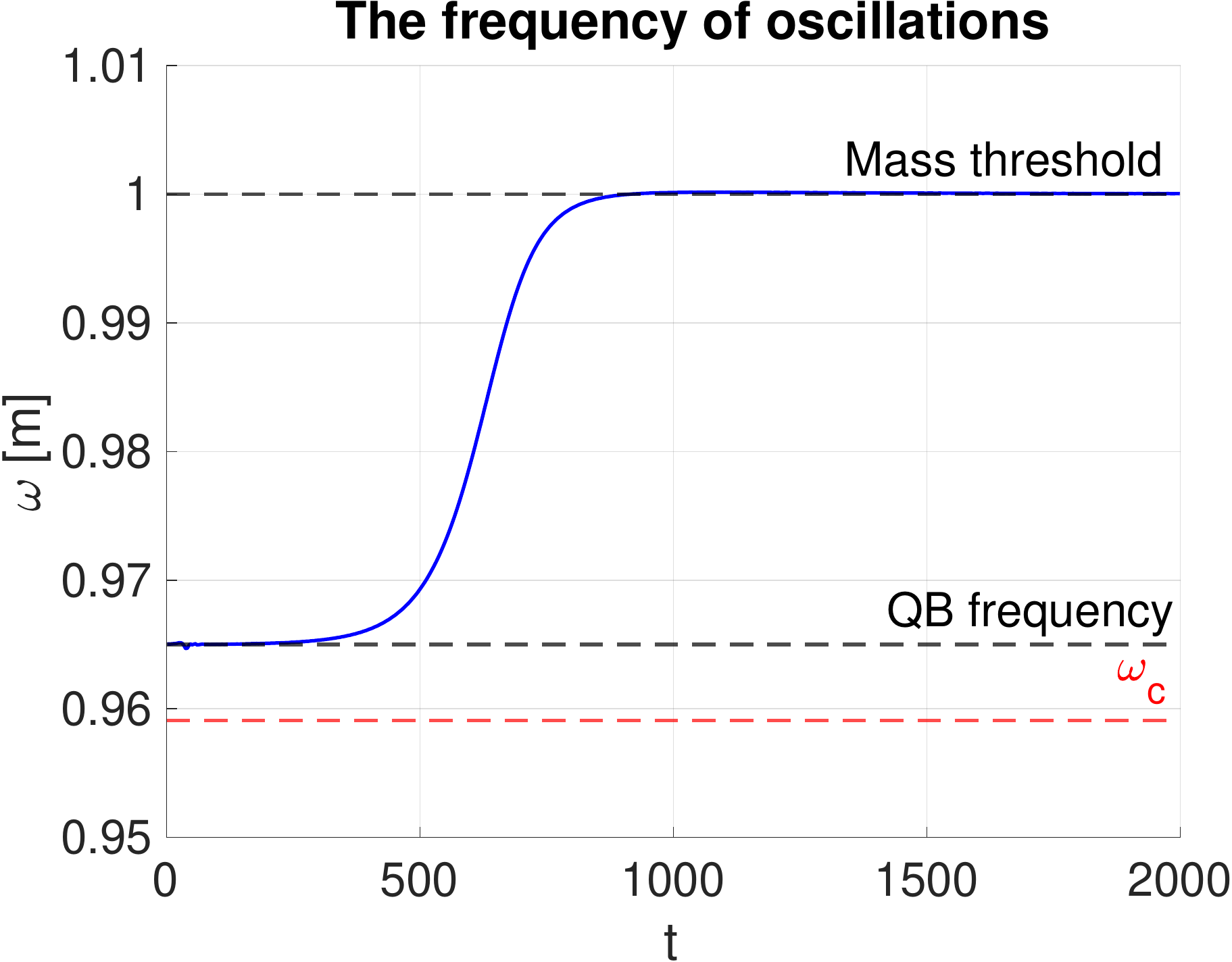}
  \caption{$D=2.25$}
\end{subfigure}%
\vspace{2mm}
\begin{subfigure}{.99\textwidth}
  \centering
  \includegraphics[width=.49\linewidth]{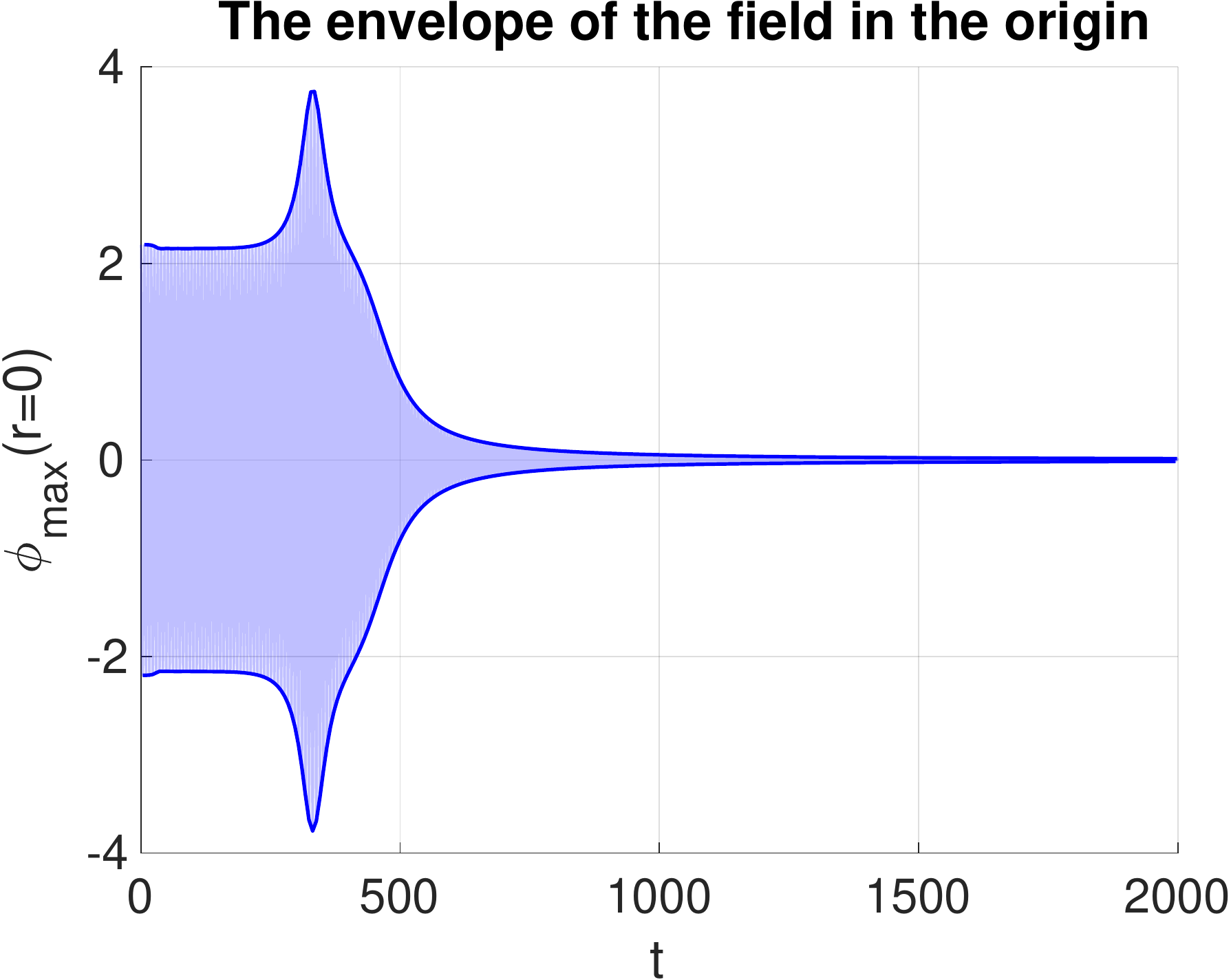}
  \includegraphics[width=.49\linewidth]{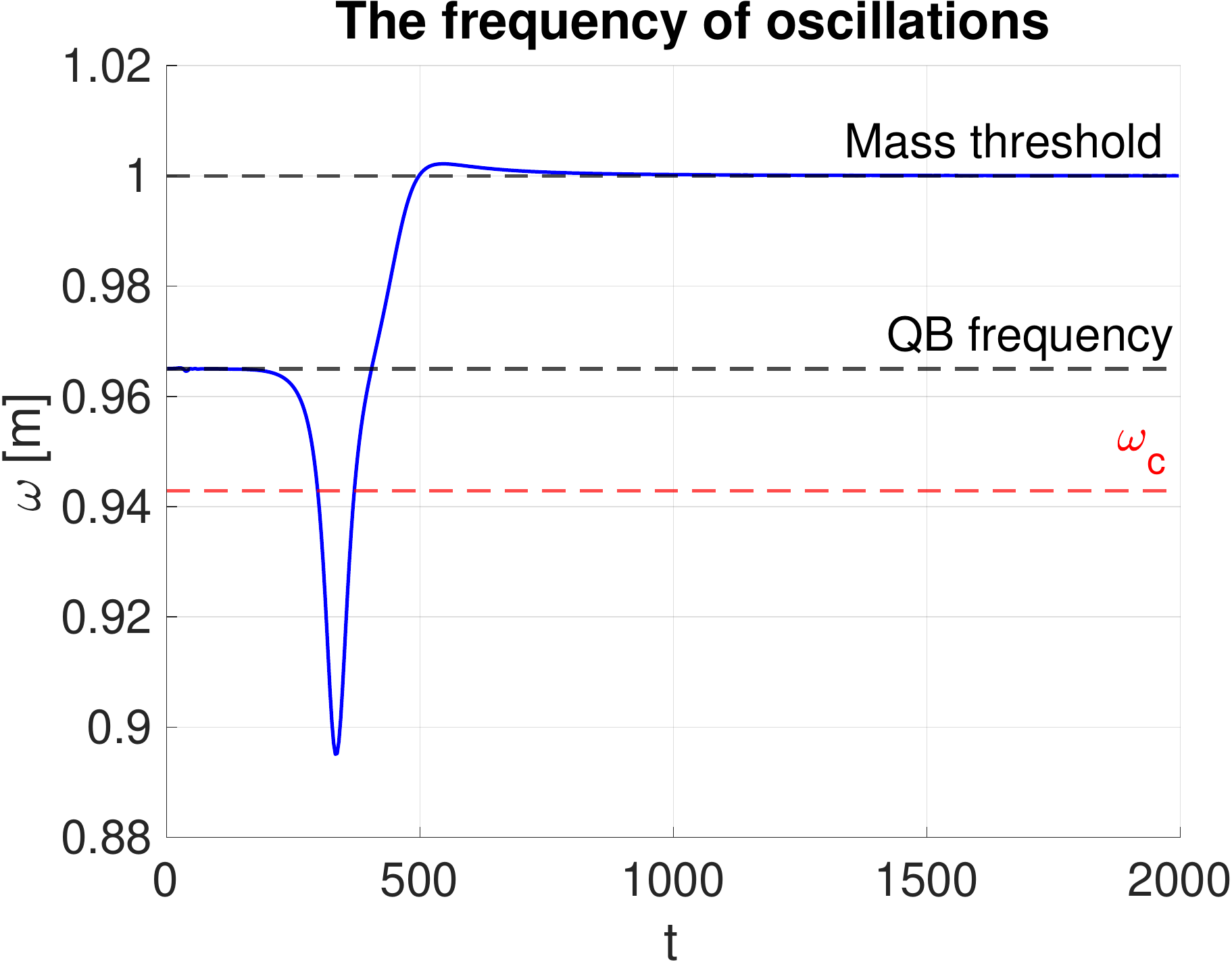}
  \caption{$D=2.50$}
\end{subfigure}
\caption{\label{decay_modes} The envelope and the frequency of the field for (a) $D=2.25$, (b) $D=2.50$, and $\omega=0.965$, $N=5$.}
\end{figure*}

We remark that for $D=1$, $E(\omega)$ is known analytically for the breather of the sine-Gordon model
\begin{equation}
    E(\omega) = 16\epsilon = 16\sqrt{1-\omega^2}\ ,
\end{equation}
which fits perfectly with our numerically obtained data.

From the curves in Fig. \ref{frequencies}. it is also apparent that the time evolution of the frequency starts at value set for the quasibreather solution, which shows that the numerical solution for the quasi-breather is of sufficiently high precision so that there is virtually no initial transient phase in the evolution corresponding to the system settling down in an oscillon configuration. Upon reaching the above obtained critical frequencies, the oscillons collapse (except in $D=2.25$, where the oscillon radiates so slowly that it was not possible to simulate long enough to reach the collapse).

In Fig. \ref{frequencies}, one can also see small oscillations in the time dependence of the frequency. These correspond to imperfections of our quasibreather solution and can be suppressed by including more harmonics when constructing the quasibreather, and also by minimizing the tail amplitude of the first radiation mode. These improvements can also increase the lifetime of the given oscillon until the collapse, but only by a small amount.

When the time evolution is started from a quasibreather solution with a frequency $\omega>\omega_c$ (i.e., above the critical frequency), the oscillon is in an unstable phase and in which it has two decay modes \cite{PhysRevD.74.124003,PhysRevD.65.084037,fodor2019review}. The resulting evolution is shown in Fig. \ref{decay_modes}. It is possible to fine tune the initial data to have the oscillon decay through a preselected mode, but we do not pursue this here.

We stress that an important outcome of these computations is that we could construct an accurate quasibreather solution, which is especially shown by the fact that the gradual evolution of frequency starts very close to the value for which the quasibreather was constructed.

\clearpage

\subsection{Low-frequency regime: Staccato decays}

\begin{figure*}
\centering
\begin{subfigure}{.49\textwidth}
  \centering
  \includegraphics[width=1\linewidth]{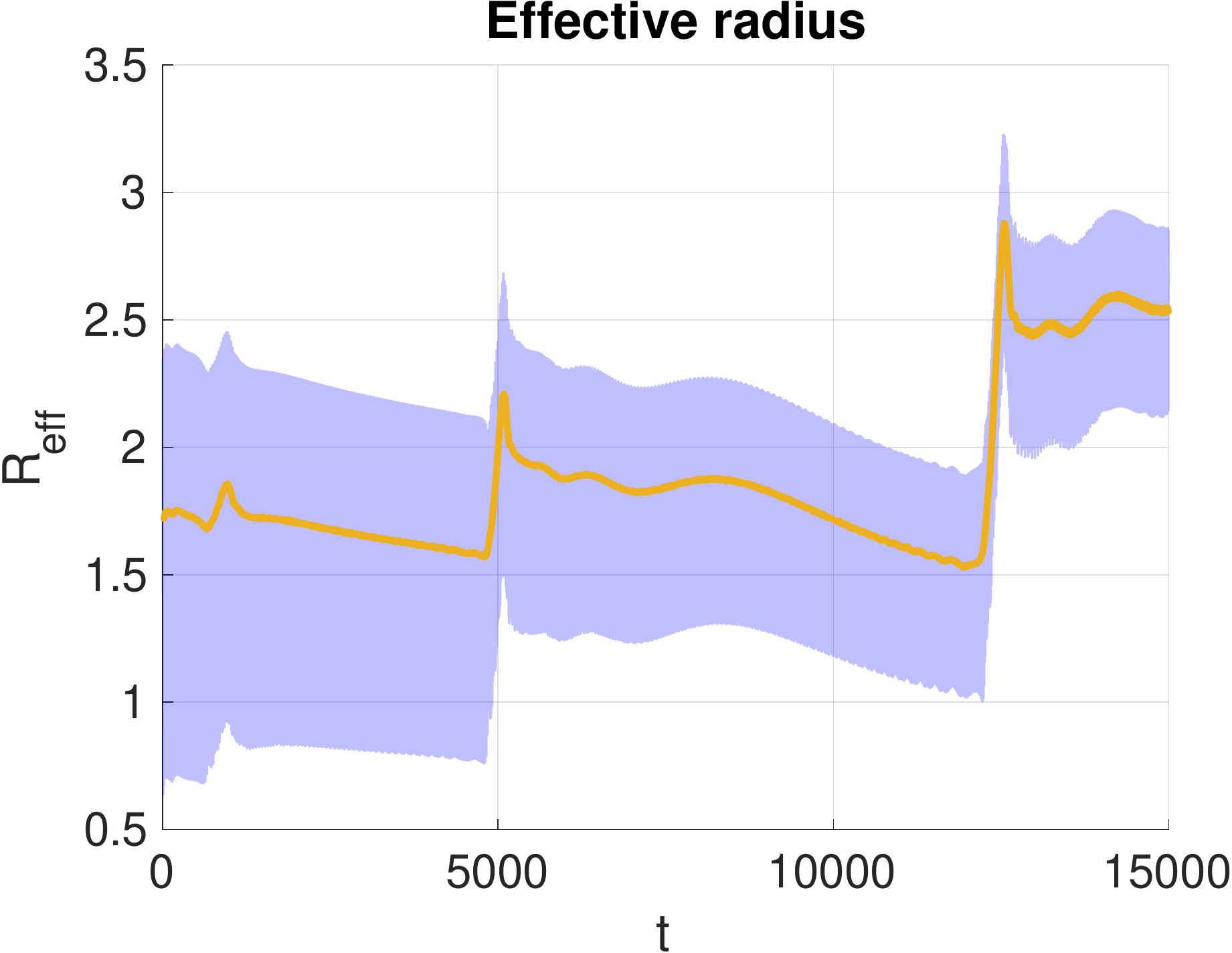}
  \caption{\label{low_freq_effrad}}
\end{subfigure}
\begin{subfigure}{.49\textwidth}
  \centering
  \includegraphics[width=1\linewidth]{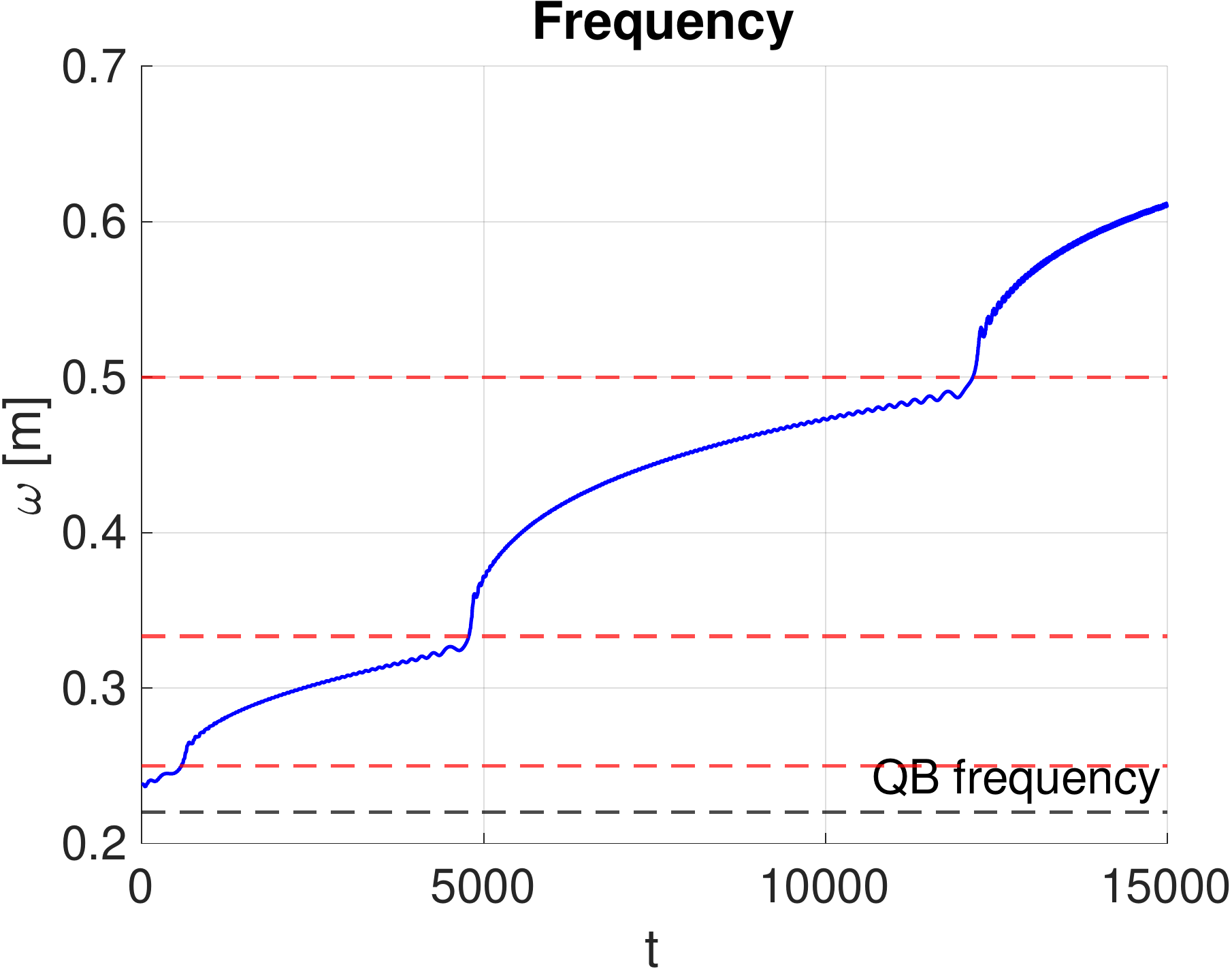}
  \caption{\label{low_freq_freq}}
\end{subfigure}
\begin{subfigure}{.49\textwidth}
  \centering
  \includegraphics[width=1\linewidth]{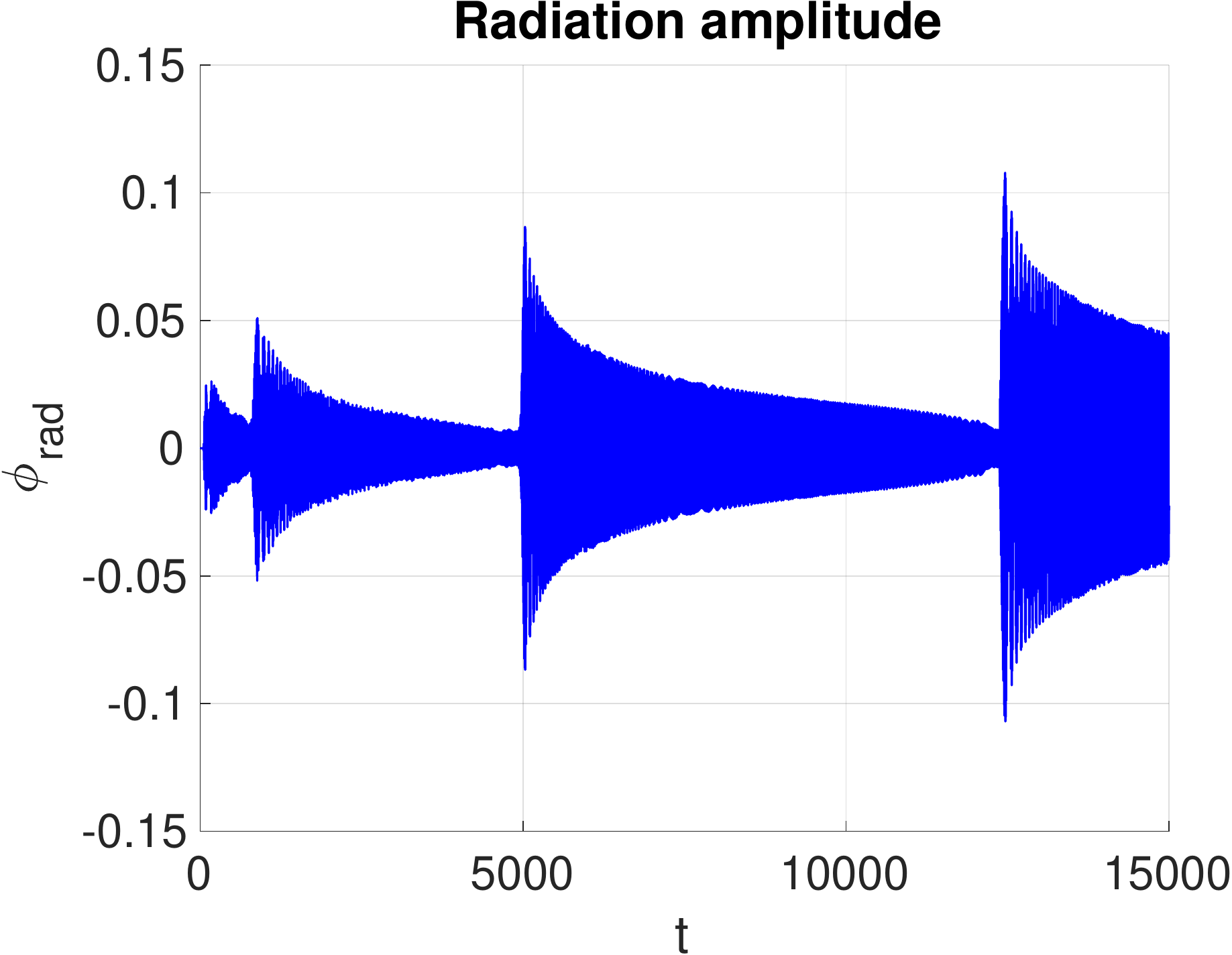}
  \caption{\label{low_freq_rad}}
\end{subfigure}
\caption{\label{low_freqd1} The time evolution of the quasibreather for $\omega=0.22$, $\lambda=0.0025$, $D=1$. The transparent curve in (a) is the full time dependence of the effective radius, the orange curve is its value averaged over a (time-dependent) period of oscillation. The sudden changes in (b) at frequencies $1/n$ (indicated by the red dashed lines, with the black dashed line showing the frequency of the initial quasibreather), and the bursts of radiation amplitude in (c) correspond to staccato steps signalling the start of the decay of a new harmonic entering the radiation continuum.}
\end{figure*}

\begin{figure*}
\centering
\begin{subfigure}{\textwidth}
  \centering
  \includegraphics[width=0.32\textwidth]{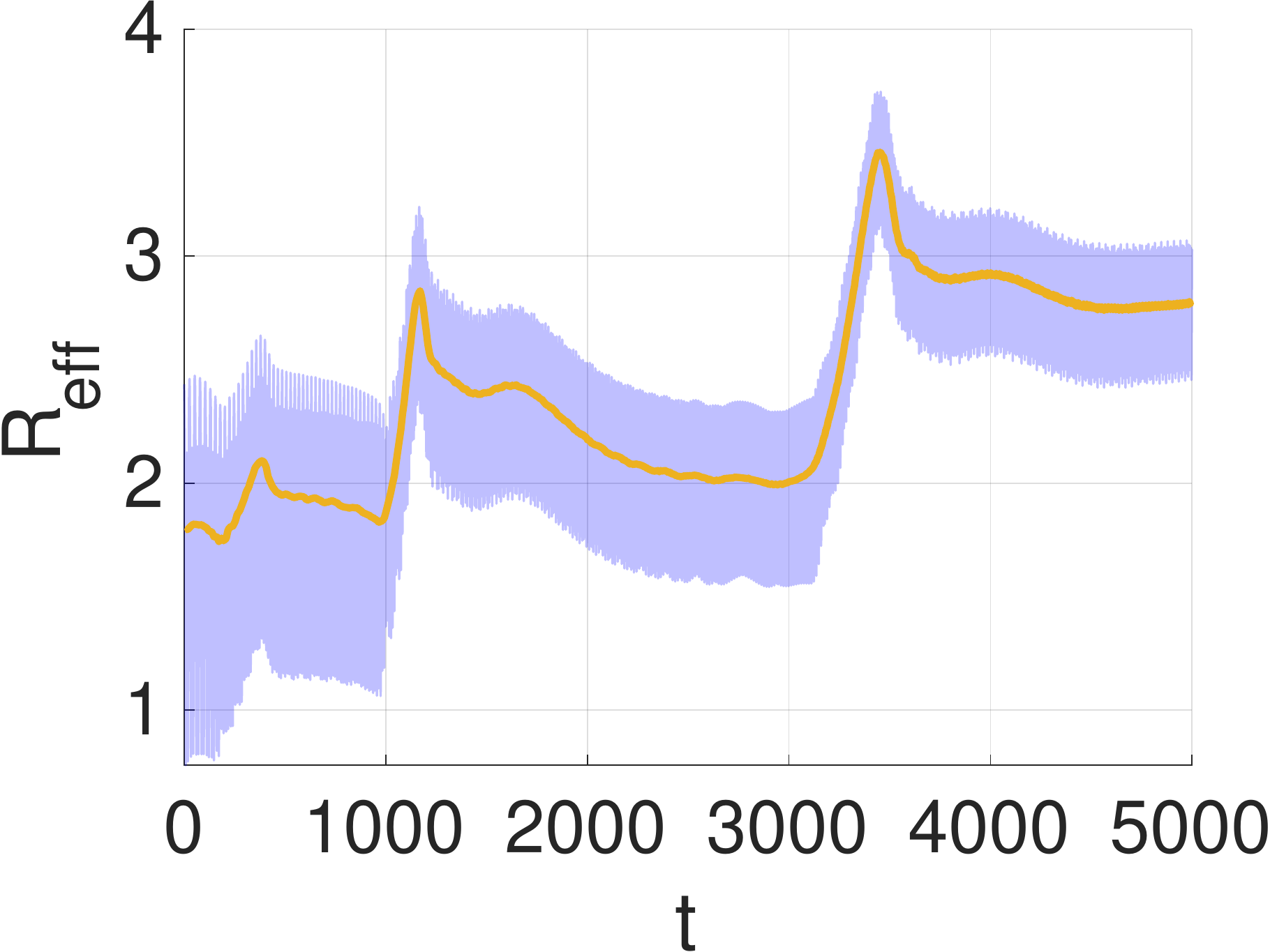}
  \includegraphics[width=0.32\textwidth]{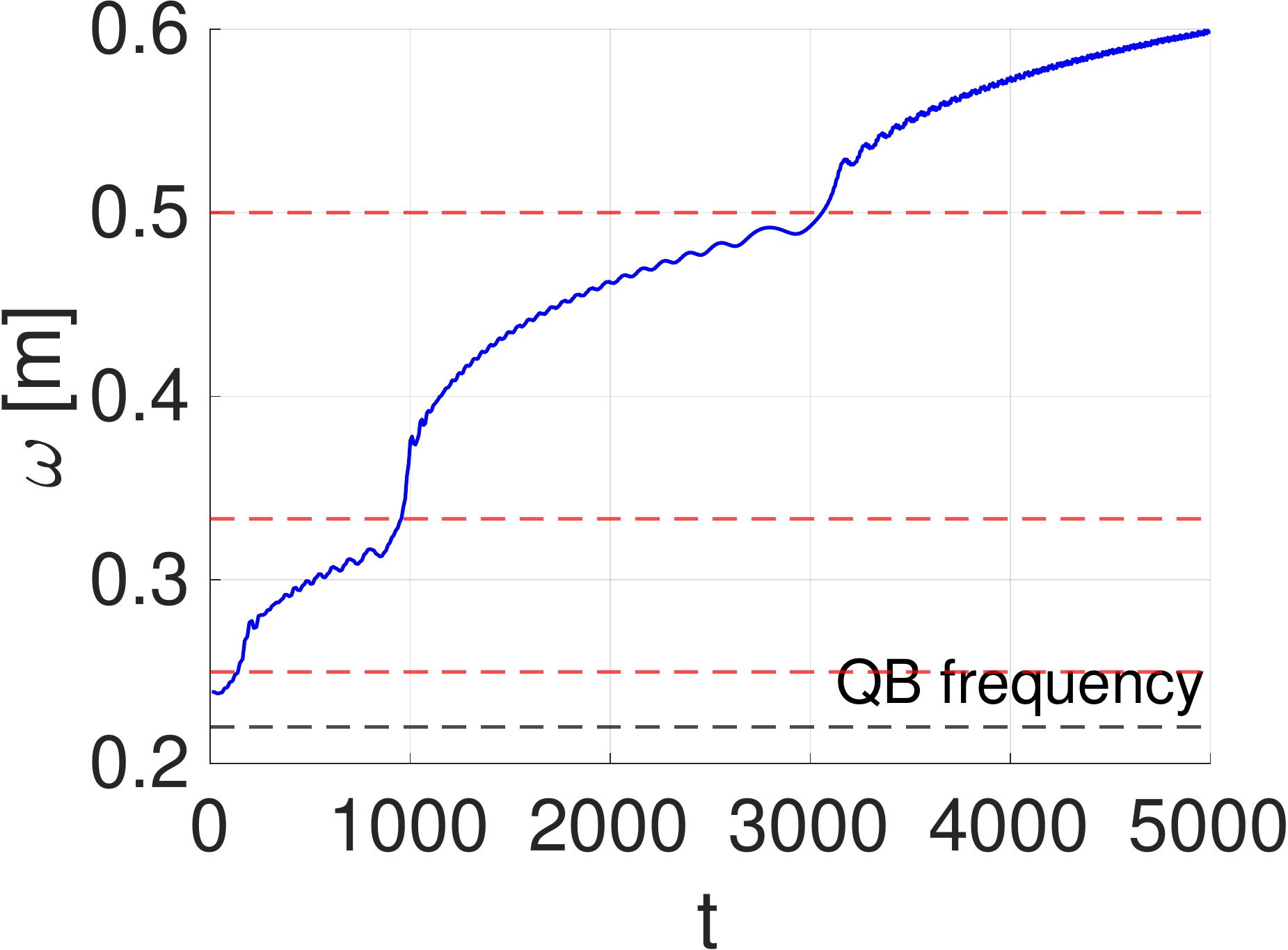}
  \includegraphics[width=0.32\textwidth]{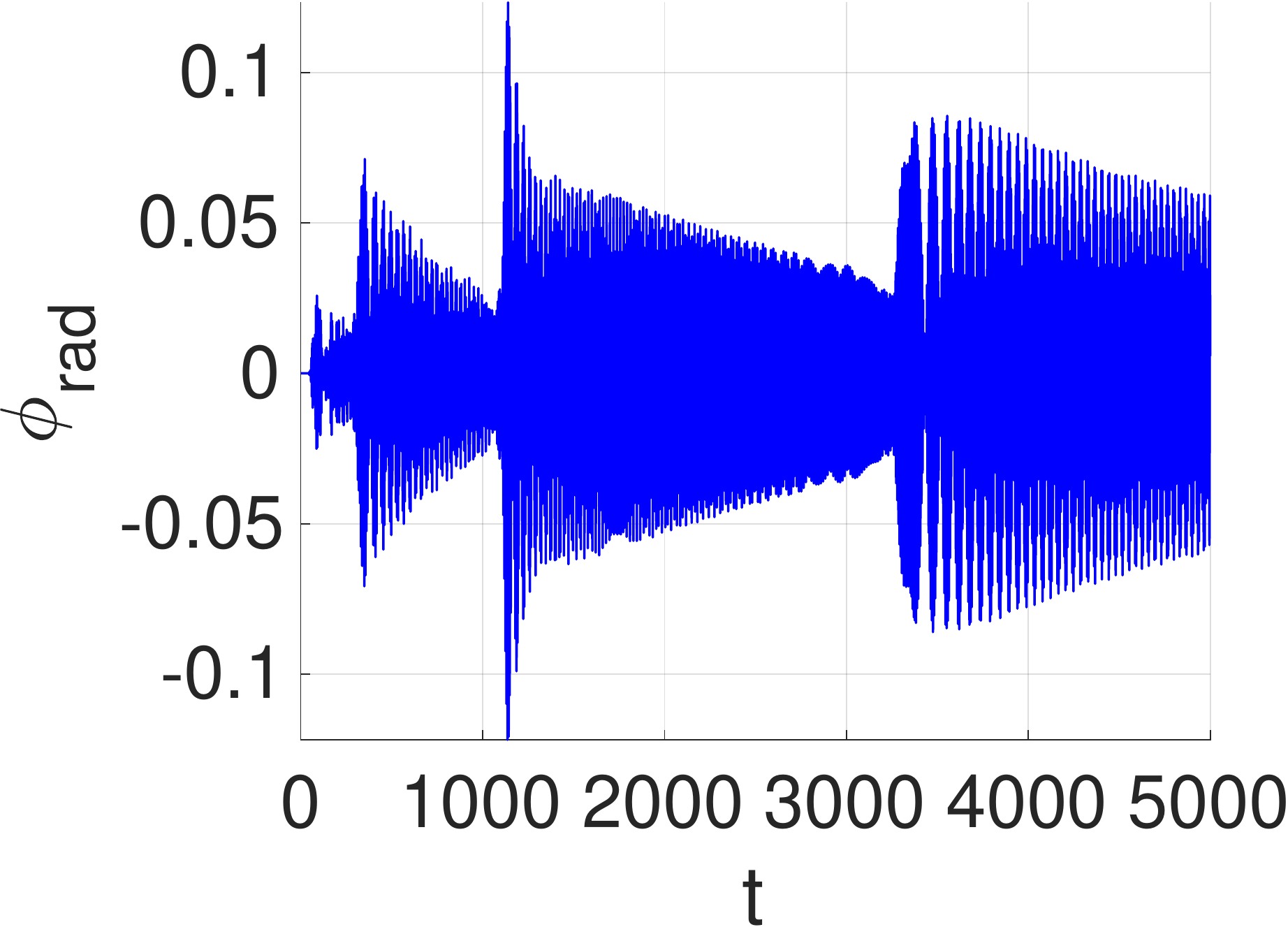}
  \caption{D=1.02}
\end{subfigure}
\begin{subfigure}{\textwidth}
  \centering
  \includegraphics[width=0.31\textwidth]{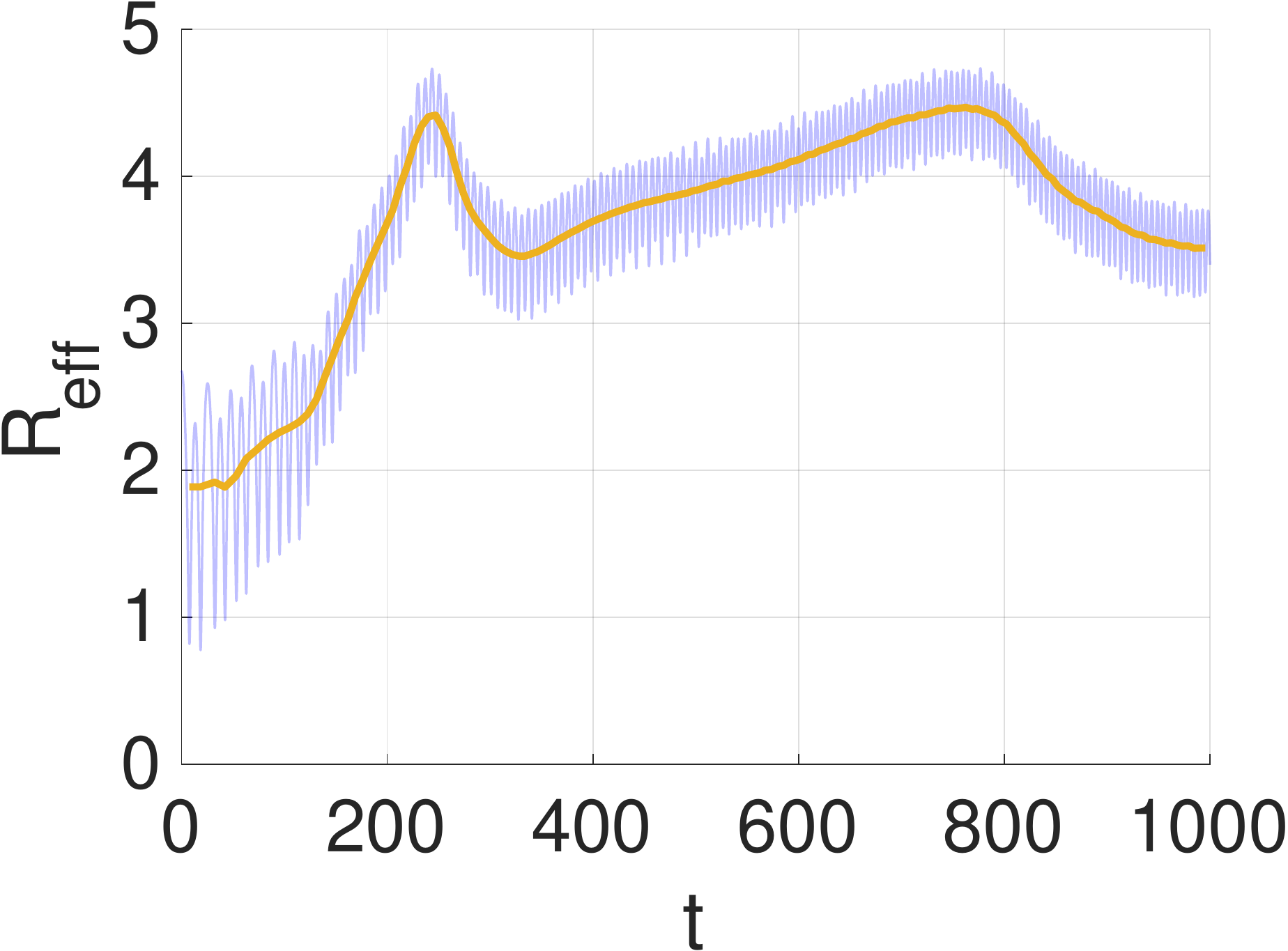}
  \includegraphics[width=0.31\textwidth]{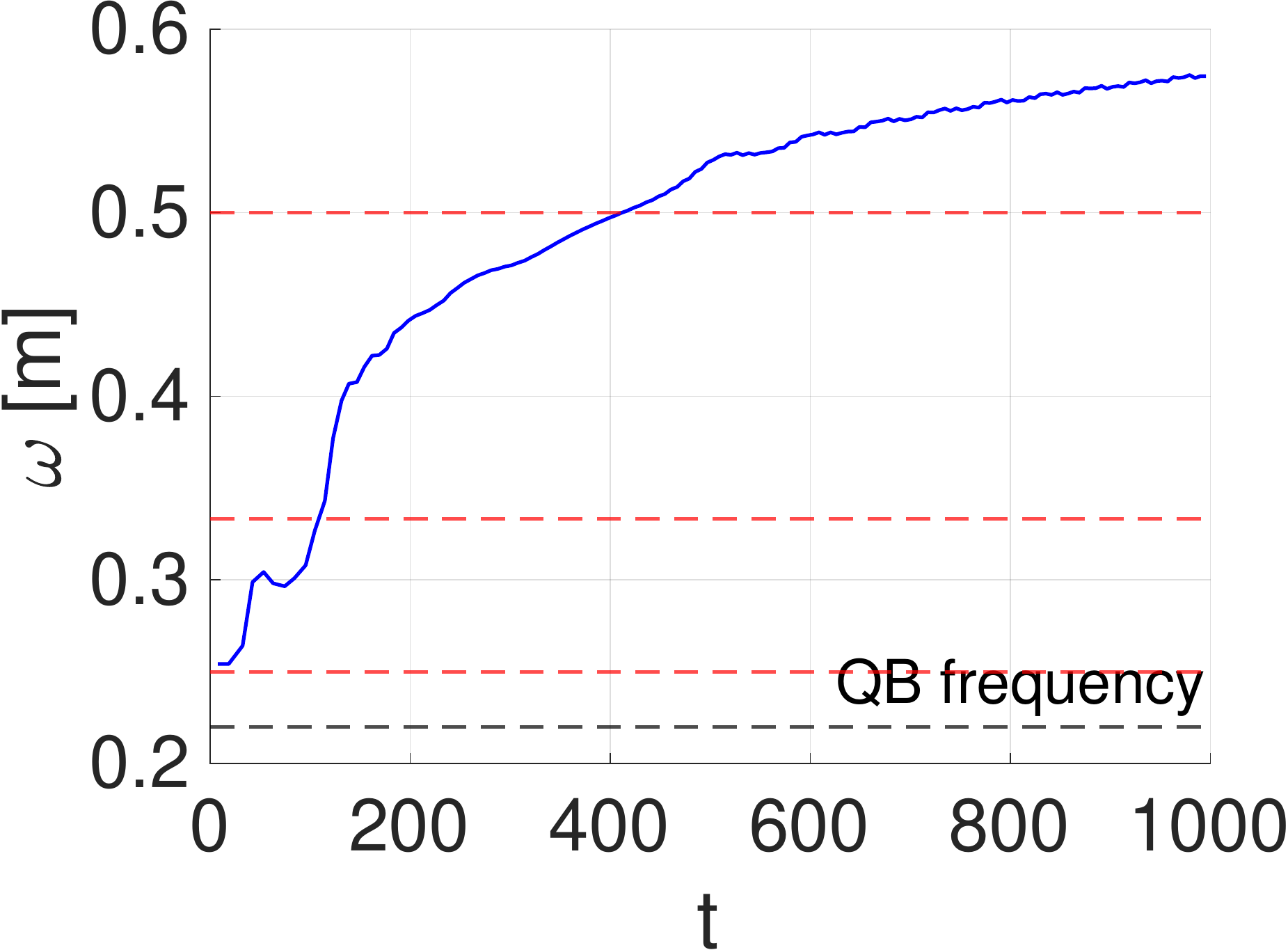}
  \includegraphics[width=0.31\textwidth]{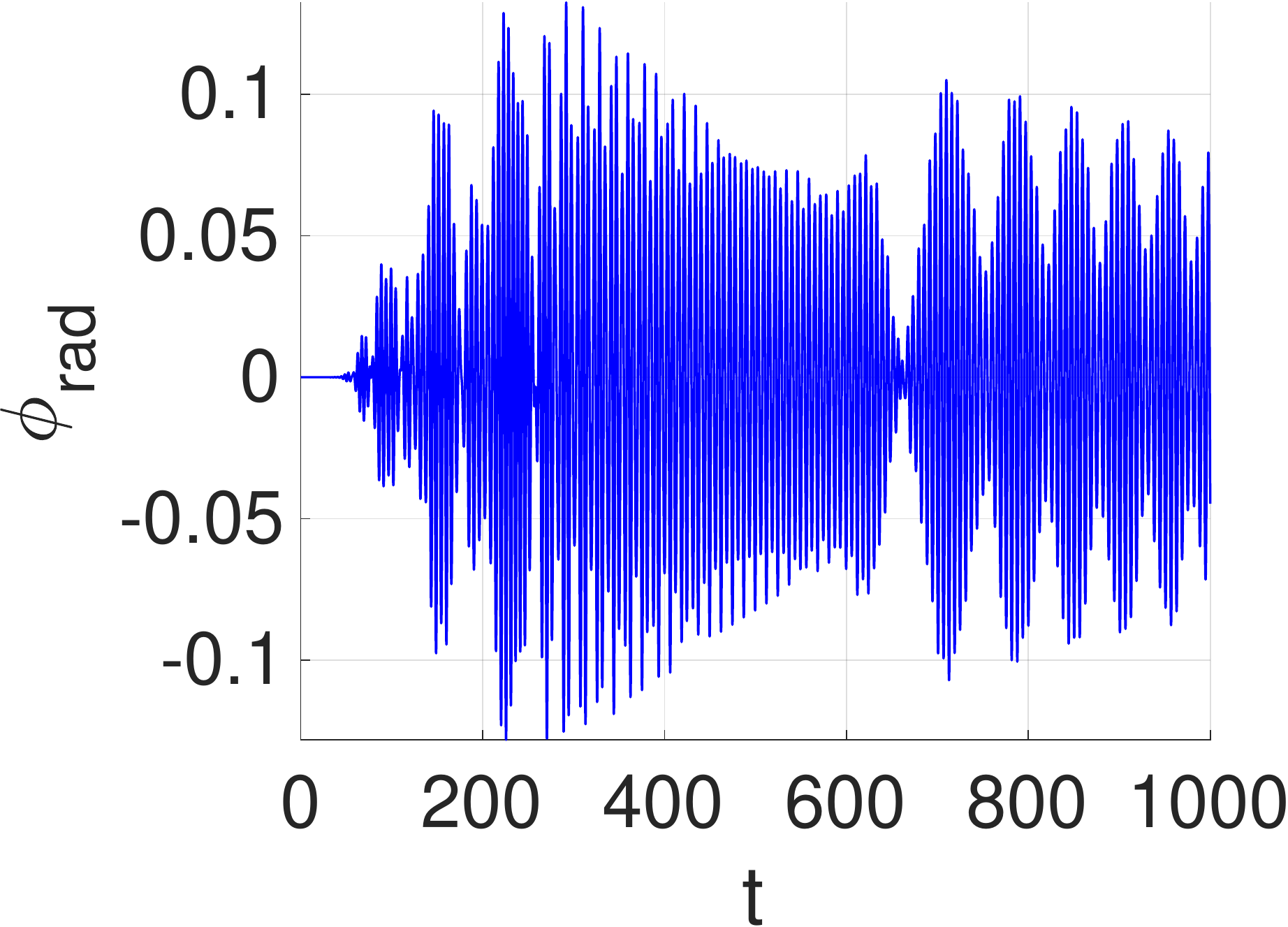}
  \caption{D=1.08}
\end{subfigure}
\begin{subfigure}{\textwidth}
  \centering
  \includegraphics[width=0.31\textwidth]{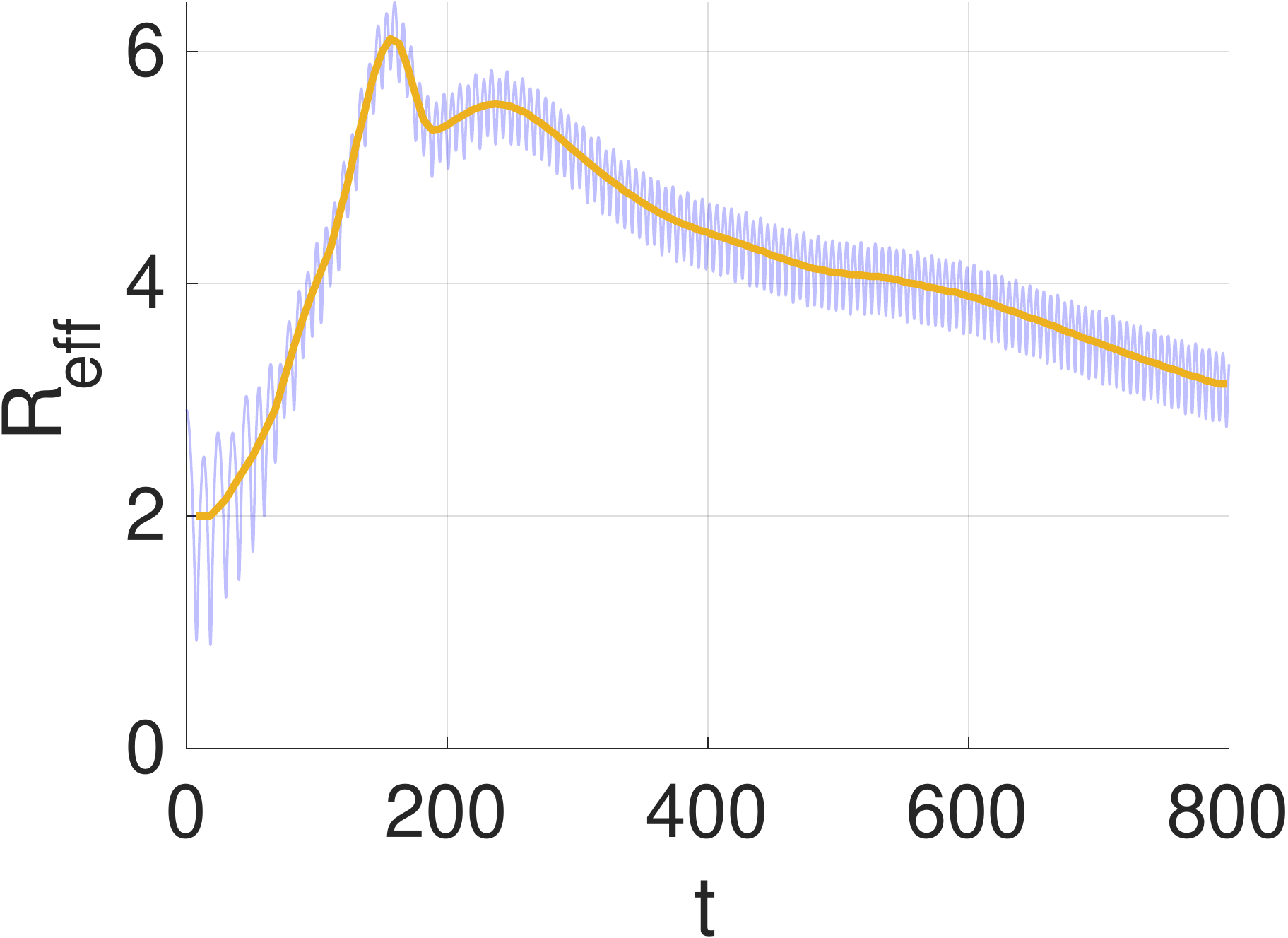}
  \includegraphics[width=0.31\textwidth]{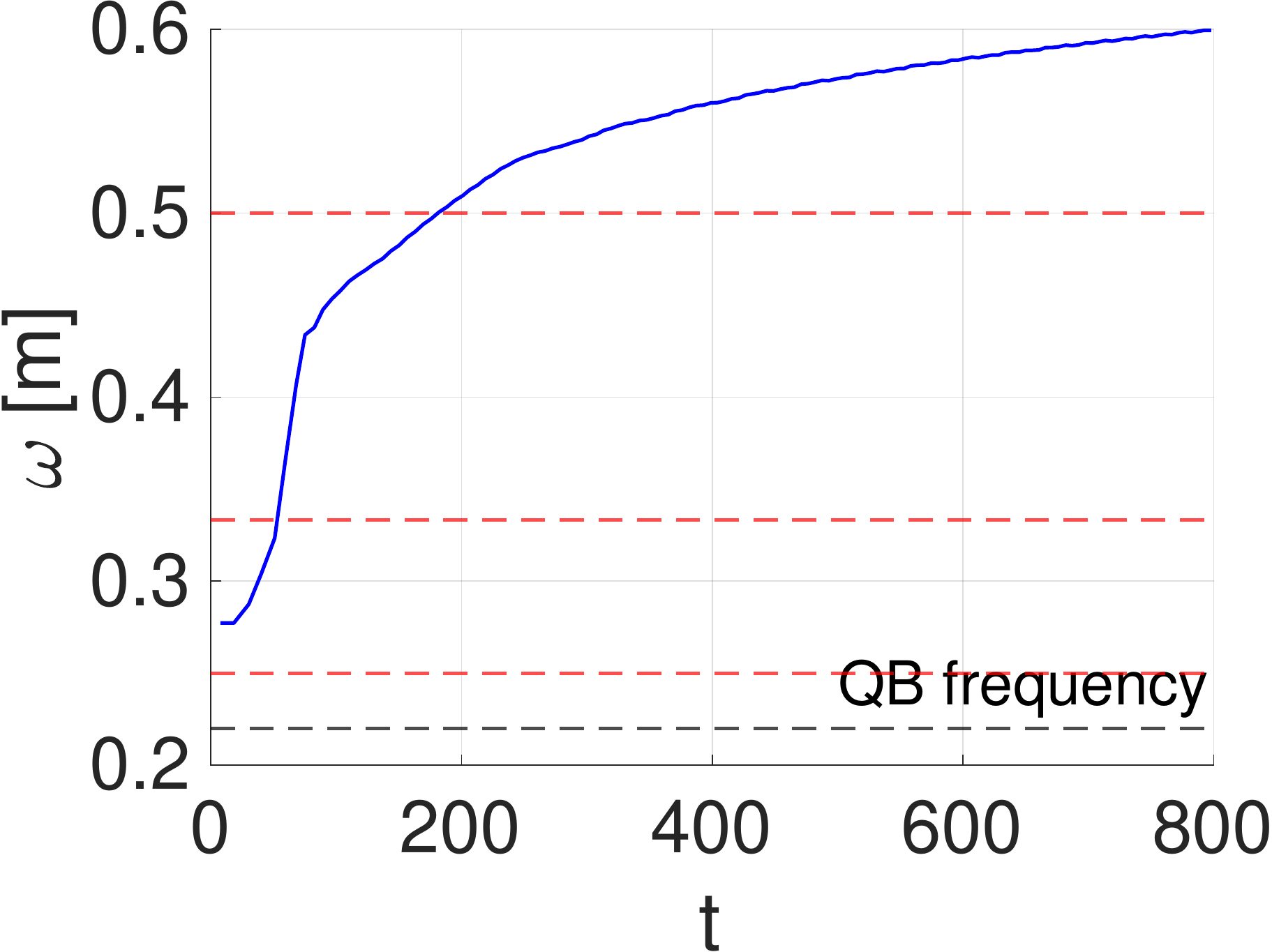}
  \includegraphics[width=0.31\textwidth]{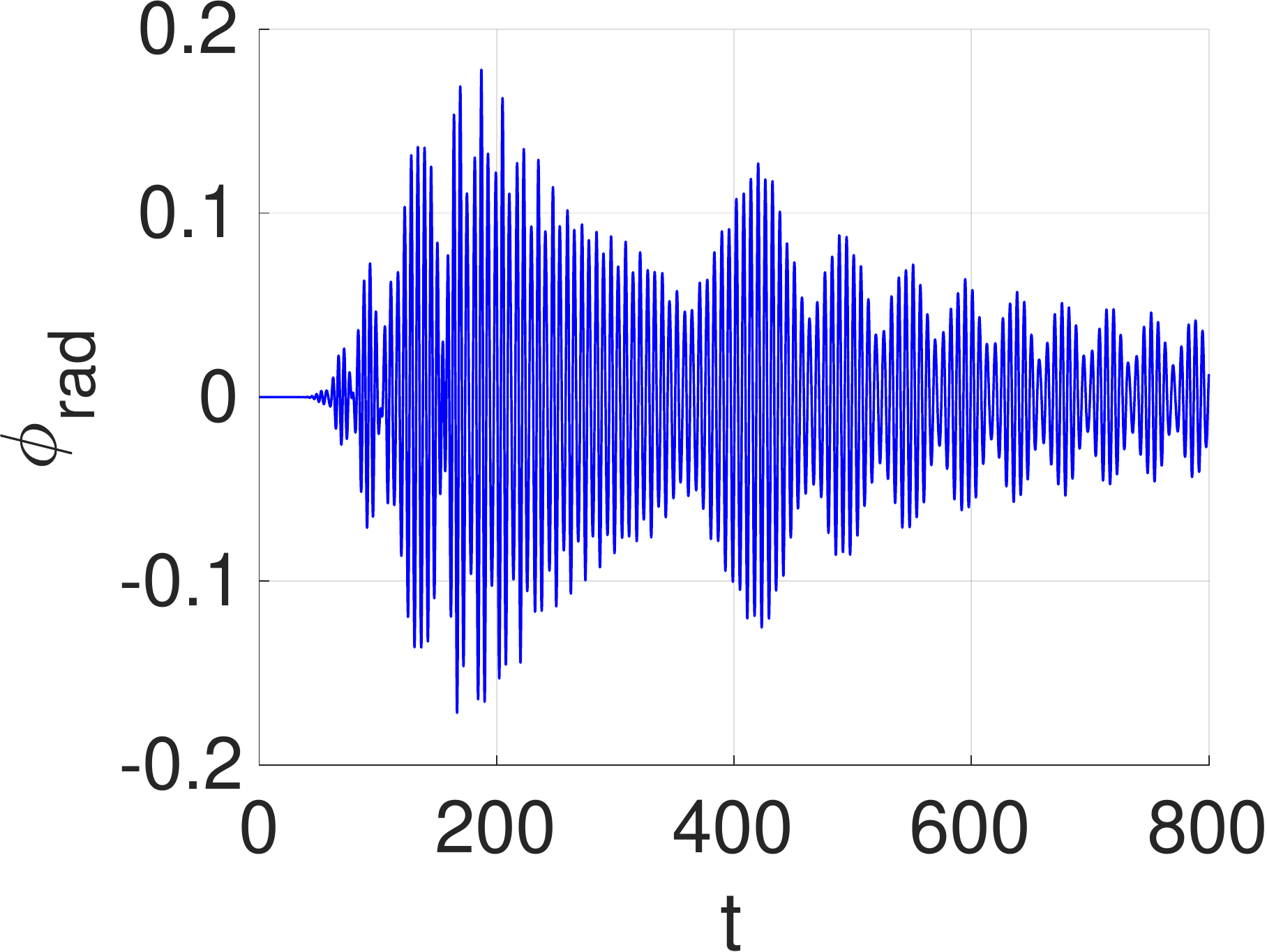}
  \caption{D=1.13}
\end{subfigure}
\begin{subfigure}{\textwidth}
  \centering
  \includegraphics[width=0.31\textwidth]{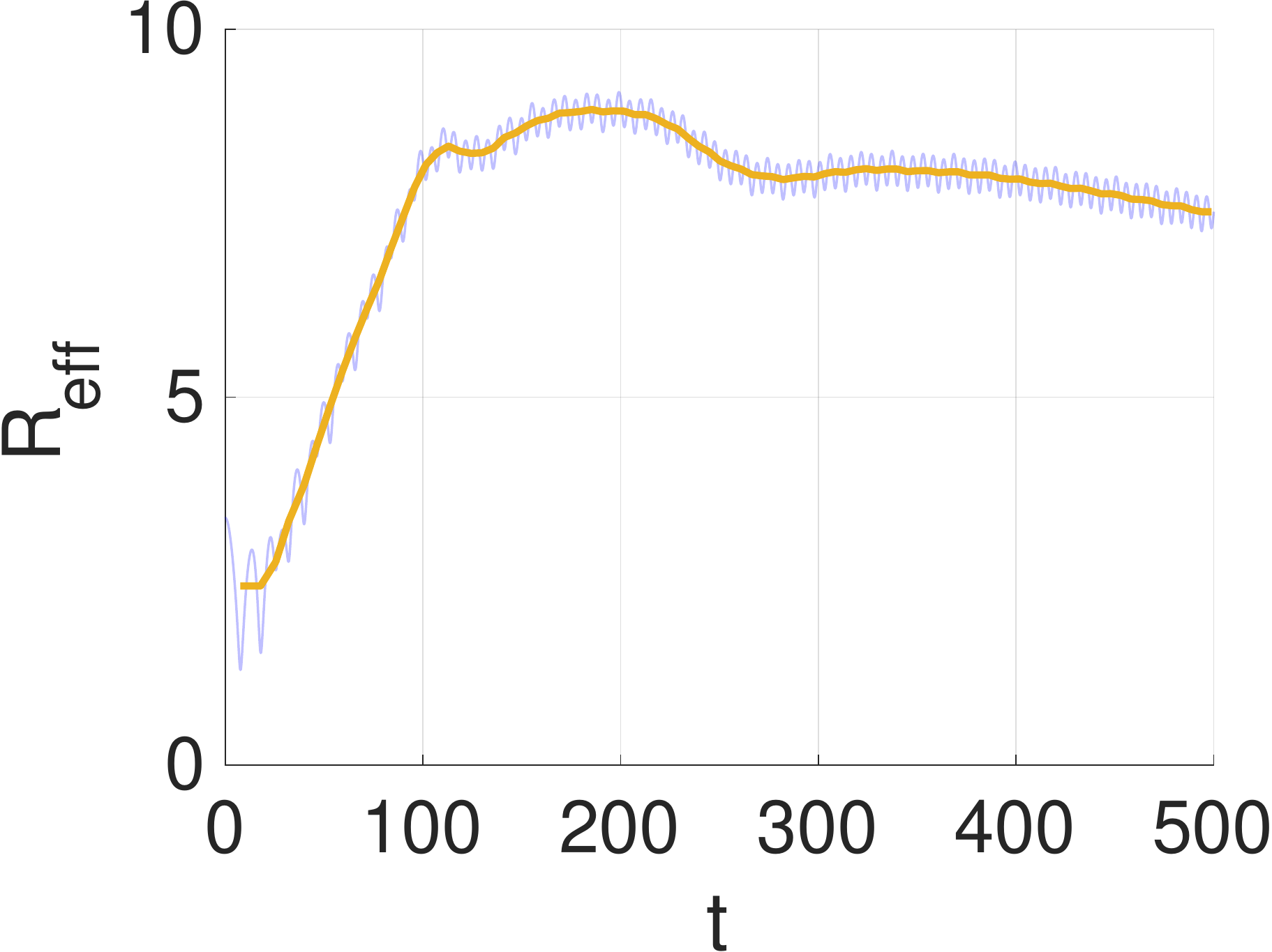}
  \includegraphics[width=0.31\textwidth]{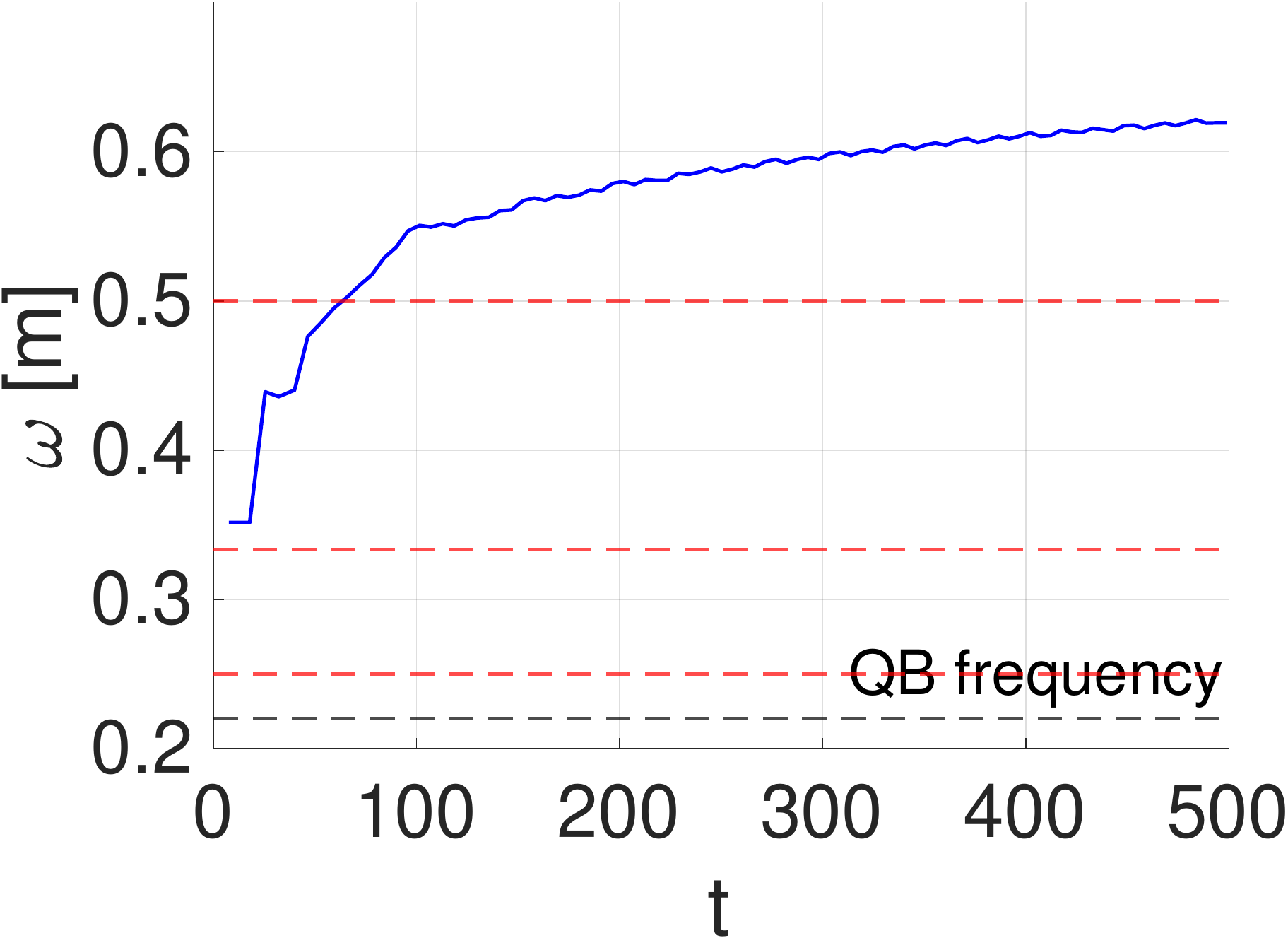}
  \includegraphics[width=0.31\textwidth]{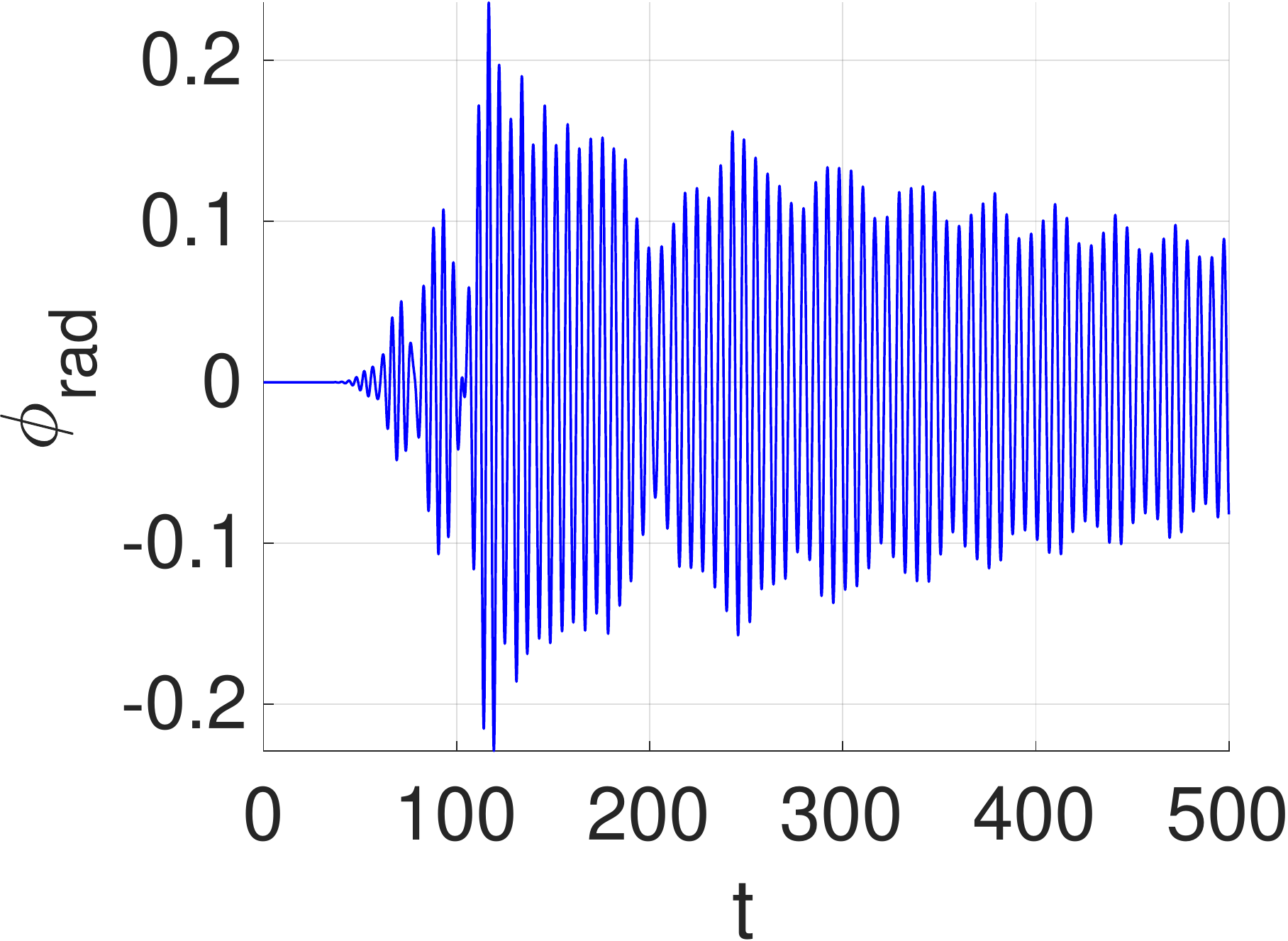}
  \caption{D=1.20}
\end{subfigure}
\caption{\label{demo} Time evolution of the effective radius, the frequency and the radiation amplitude of the $\omega=0.22$ quasibreather for $\lambda=0.0025$ in different dimensions, with the same conventions as in Fig. \ref{low_freqd1}. Notice the large variation of the time scales with $D$. In $D=1.02$, the staccato steps are clearly visible; for $D=1.08$ and $D=1.13$ they are still recognisable, but for $D=1.13$ the $n=4$ step already seems to disappear. For $D=1.20$, the characteristics of staccato decay are hardly recognisable anymore, but a sudden increase in the radiation field is still visible when the frequency rises above $\omega=1/2$.}
\end{figure*}

\begin{figure*}
\centering
\begin{subfigure}{.49\textwidth}
  \centering
  \includegraphics[width=1\linewidth]{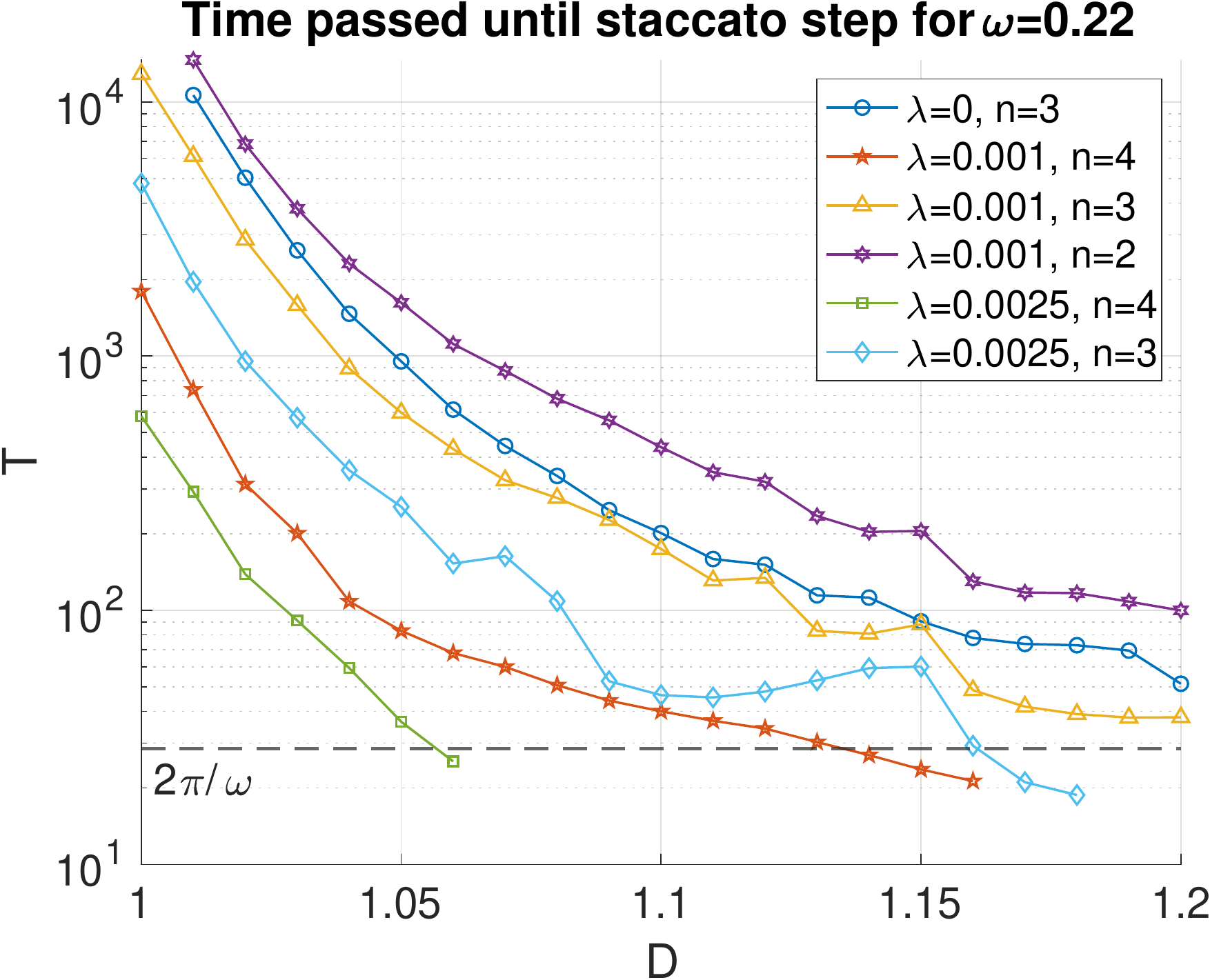}
  \caption{}
\end{subfigure}%
\begin{subfigure}{.49\textwidth}
  \centering
  \includegraphics[width=1\linewidth]{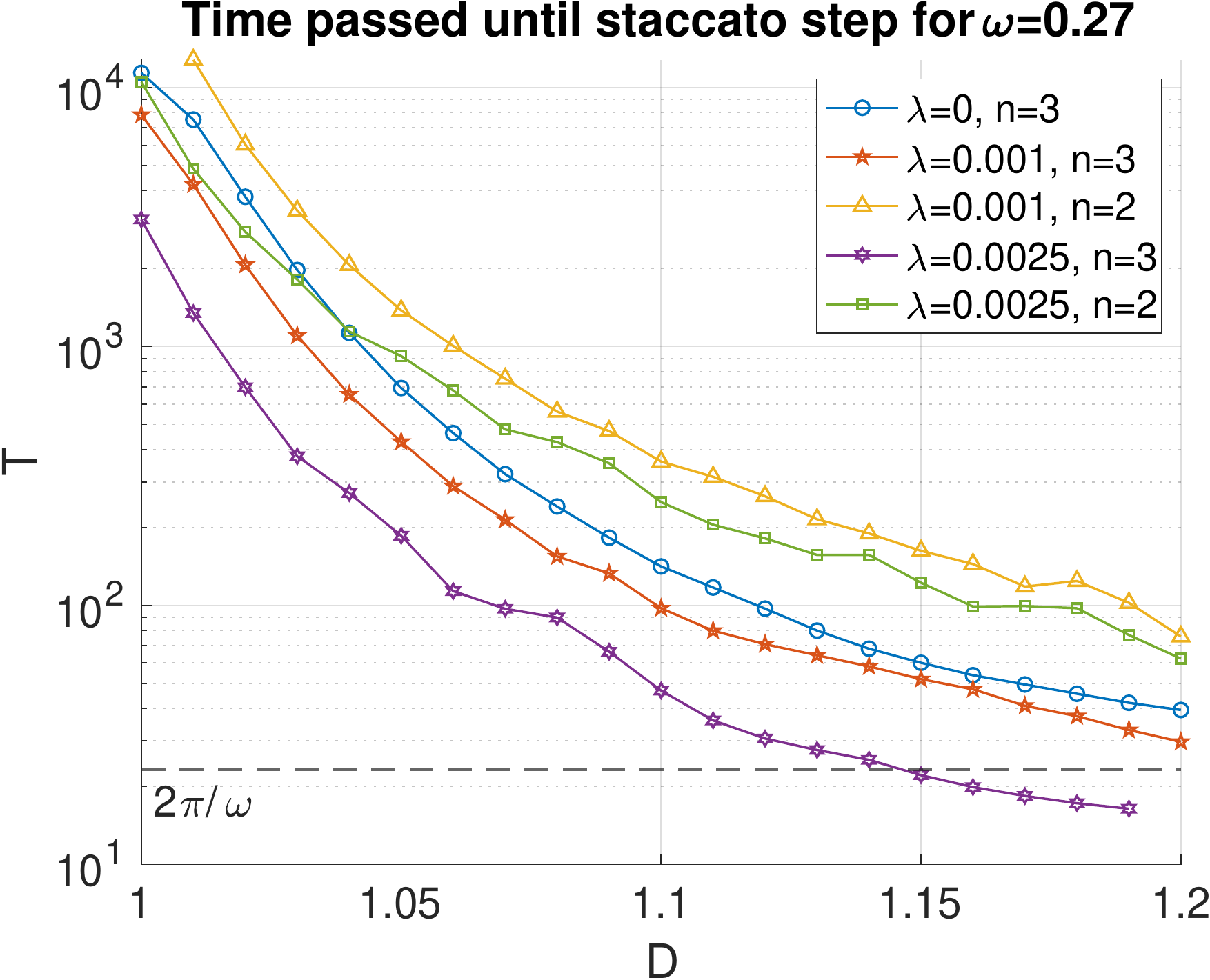}
  \caption{}
\end{subfigure}
\begin{subfigure}{.49\textwidth}
  \centering
  \includegraphics[width=1\linewidth]{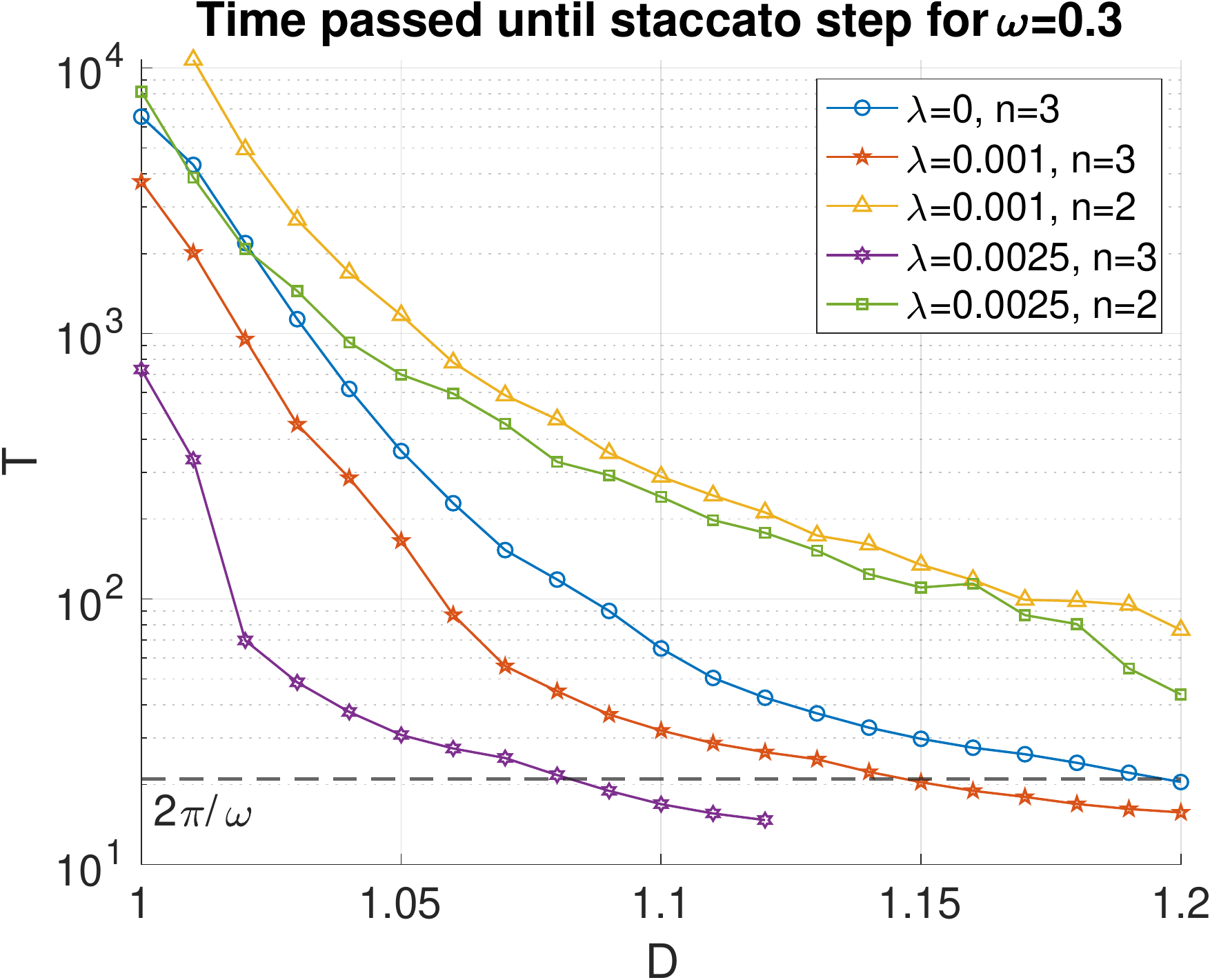}
  \caption{}
\end{subfigure}
\caption{\label{low_freq_T_D} Time elapsed for a staccato step corresponding to the $n$th harmonic for different frequency quasibreather initial profiles. When $\lambda$=0, even harmonics are all zero in Eq. (\ref{Ansatz}), so there is no staccato step for these modes. A horizontal line at $2\pi/\omega$ denotes one time period. Below this line, there is no chance of seeing the traces of staccato decay, as they fade into the transients of the early time evolution. When a curve ends prematurely it signals that the given staccato step is not visible anymore in higher dimensions. }
\end{figure*}

In this section we present our results for low-frequency quasibreathers, i.e. ones with frequency $\omega<m/2$. Our aim is to study the staccato decay recently discovered in in $D=1$ \cite{Dorey_2020}, in higher spatial dimensions $D>1$. 

The reason for the staccato decay is that the time evolving frequency $\omega(t)$ crosses a threshold value $1/n$ with $n$ a positive integer, corresponding to the sudden release of the energy stored in the $n$th harmonic into radiation. This manifests in short outbursts of radiation that can be measured by the amplitude of the field at a position far from the core, as well as in quick jumps in frequency due to the sudden energy loss and also in a temporary increase of the effective radius corresponding to the outflow of energy from the core. These signals can be seen in Fig. \ref{low_freqd1}, where we used $\lambda\neq 0$ in order to have both odd and even harmonics, and therefore the corresponding staccato bursts, present in the time evolution.

Since the mechanism described above seems generic enough, staccato decays are expected to occur in higher dimensions as well; however it turned out that our simulations showed no trace of them for $D\gtrsim 1.5$. To investigate closer, we constructed quasibreather profiles with $\omega=0.3$, $0.27$, $0.22$, for different $\lambda$'s for spatial dimensions $D \in [1, 1.20]$, and simulated their time evolution. It turns out that staccato steps can indeed be seen for low enough values of $D$, as shown in Fig. \ref{demo} for $\lambda=0.0025$. For this simulation we used only odd harmonics in the ansatz (\ref{Ansatz}), even though for $\lambda \neq 0$ the potential is not even. This resulted in gains in computing time; however, due to the less accurate quasibreather profiles now we can see that the time evolution of the oscillon starts from a frequency higher than the one for which the quasibreather was constructed. It is apparent that with increasing $D$ the dynamics accelerate, with the staccato steps moving to earlier times until they eventually fade into the transient region that takes place at the early stages of the time evolution when the initial profile relaxes to the nearest oscillon profile.

For a more detailed understanding we determined the time it takes until the appearance of the staccato step corresponding to the $n$th harmonic, for three different initial quasibreather frequencies and values of $\lambda$, as a function of the number of spatial dimensions $D$. The results, summarized in Fig. \ref{low_freq_T_D} show that the time intervals shrink fast when $D$ is increased, and very soon drop below the period of the quasibreather itself, which means that they happen in such a quick succession that they cease to make sense since the change of frequency cannot be defined on time scales shorter than the period of the oscillon itself. What happens is that the staccato steps become indistinguishable from the short transient period at the start of the time evolution of the oscillon itself. We remark that the acceleration of the oscillon decay dynamics with increasing number of spatial dimensions is also consistent with the behaviour observed in the high frequency regime, c.f. the results shown in Fig. \ref{frequencies}.

\section{Conclusions}\label{sec:conclusions}

In the present work we studied the dependence of oscillon decay on the number of spatial dimensions. Using an improvement of a previous method we constructed accurate quasibreather solutions to use them as initial conditions for time evolution in the $(D+1)$-dimensional sine-Gordon theory. The method turned out to work especially well in the high-frequency regime, where we computed the energy-frequency curve of quasibreathers and demonstrated the existence of a minimum at some critical frequency $\omega_c<1$, previously inferred using the small-amplitude expansion. By computing the time evolution we demonstrated that this critical frequency is exactly the value which determines the condition of the sudden collapse of the oscillon. We also demonstrated that this minimum of the energy-frequency curve disappears for $D\leq 2$, again in accordance with arguments from the small-amplitude expansion.

In the second part of our investigations we considered the time evolution of low frequency quasibreathers in sine-Gordon model and its deformation by a $\phi^4$ interaction, to investigate whether the staccato decay mechanism observed in \cite{Dorey_2020} exists in dimensions $D>1$. Staccato decay is a robust feature in one spatial dimension, and also appears in other theories such as $\phi^6$ and hyperbolic $\phi^4$ models \cite{Dorey_2020}, even when the time evolution does not start from a finely tuned quasibreather, but instead from an oscillon which emerges from a kink-antikink scattering. In our study, we found that the staccato steps accelerate with increasing $D$, and in fact already at $D=1.2$ all the staccato steps take place so fast that their characteristic signals are no longer discernible from the transients of the early dynamics. As a result, staccato bursts are not observable in physically relevant (integer) number of spatial of dimensions larger than one, at least for the potentials considered here.

One possible way out is to find a fine-tuned field theoretical potential for which the oscillon decay is slow enough so that the time interval between staccato bursts are longer than the oscillon period. Indeed it is possible to find "islands of longevity" by fine-tuning potentials \cite{Cyncynates_2021}, however all known cases have frequencies $\omega>m/2$ for which staccato decay does not exist.

Moreover, we suspect that such fine tuning is extremely difficult, if not outright impossible, based on the following intuitive argument. The decay of oscillons can be viewed as a feature resulting from the nonintegrability of theory. For integrable theories such as the $D=1$ sine-Gordon model the localised periodic solutions are exactly stable, corresponding to the breathers \footnote{This is the reason why, in order to see the staccato decay in the one-dimensional sine-Gordon model it is necessary to switch on an integrability breaking coupling $\lambda$}. Increasing the dimension $D$ introduces another source of integrability breaking, which is present even in the absence of an explicit integrability breaking coupling. Our numerics indicates that this additional breaking of integrability leads to the swift acceleration and disappearance of staccato decay for $D\gtrsim 1$, not only for the sine-Gordon deformed with $\phi^4$ considered in \cite{Dorey_2020}, but even for the pure sine-Gordon case which for $D=1$ is integrable. Starting from a (numerically constructed) quasibreather configuration helps observing the staccato steps by suppressing the initial transient responsible for the sudden frequency increase at the beginning of the time evolution, but even so the steps become indistinct well before reaching $D=2$.

To sum up, the fascinating staccato decay of oscillons is likely to be confined to one spatial dimension, at least for simple potentials like the ones we considered in this work. For more complex potentials, the lifetimes and the decay rates can have very complicated dependence on the parameters, so the possibility of staccato decay in higher dimensions cannot be excluded with certainty, and deserves further investigation.


\begin{acknowledgments}
We thank P. Dorey for useful comments on the manuscript. 
This work was partially supported by the National Research, Development
and Innovation Office (NKFIH) under the research Grant K-16 No. 119204,
and also by the Fund TKP2020 IES (Grant No. BME-IE-NAT), under the
auspices of the Ministry for Innovation and Technology. G.T. was
also supported by the the National Research, Development and Innovation
Office (NKFIH) via the Hungarian Quantum Technology National Excellence
Program, Project No. 2017-1.2.1-NKP- 2017-00001. B.N. was also supported by the {\'U}NKP-20-2 New National Excellence Program of the Ministry for Innovation and Technology from the source of the National Research, Development and Innovation Fund.
\begin{center}
        \includegraphics[width=0.2\textwidth]{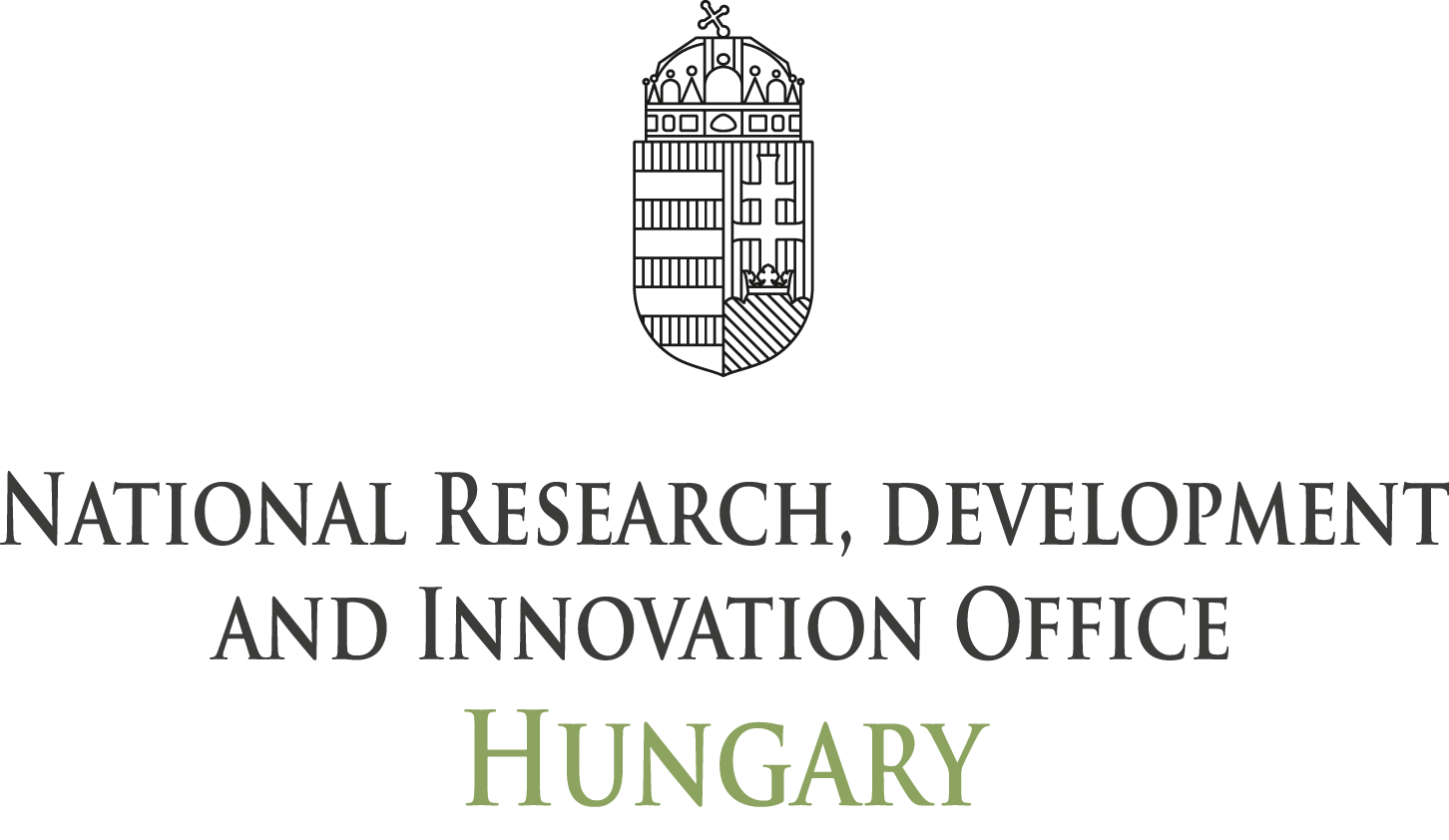}
        \includegraphics[width=0.2\textwidth]{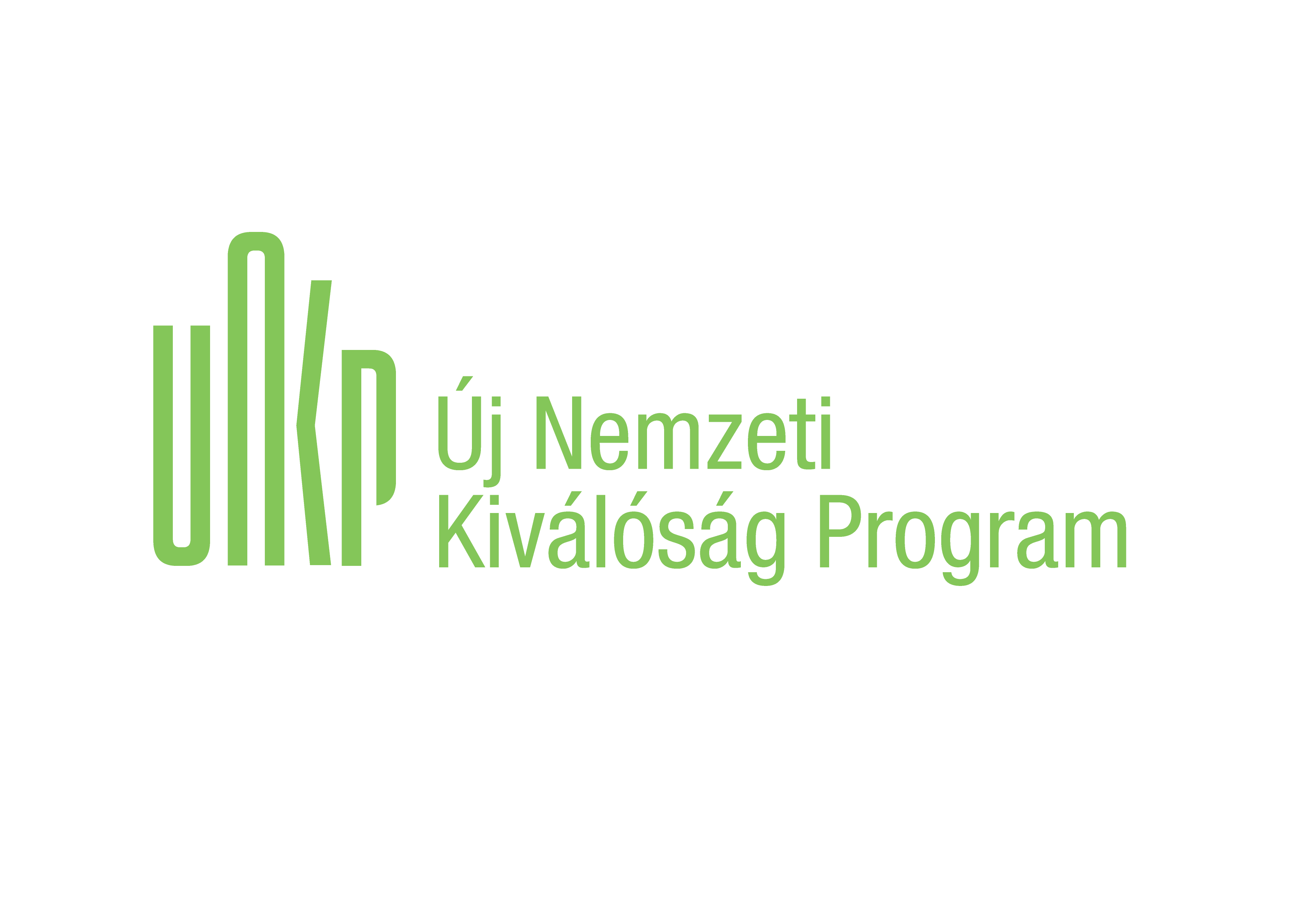}
\end{center}
\end{acknowledgments}

\bibliographystyle{utphys}
\bibliography{oscillon}
\appendix
\section{Energy of a quasibreather}
\label{QBenergy}

To evaluate the energy density (\ref{energy_density}), it is necessary to calculate the time and spatial derivatives of the quasi-breather given by (\ref{Ansatz})
\begin{eqnarray}
    &&\frac{\partial \phi}{\partial t} = \sum_{m=1}^N \phi_m(r) (-m\omega) \sin(m\omega t)\ ,\nonumber\\ 
    &&\frac{\partial \phi}{\partial r} = \sum_{m=1}^N \frac{\partial \phi_m(r)}{\partial r} \cos(m\omega t)\ .
    \label{deriv}
\end{eqnarray}
Due to the truncation at some finite $N$, the quasibreather constructed from the ansatz (\ref{Ansatz}) is not an exact solution of the equation of motion (\ref{EOM}), and therefore the energy functional evaluated with the quasibreather ansatz (\ref{Ansatz}) oscillates in time. This can be eliminated by averaging over a period as follows:
\begin{equation}
\begin{split}
    \langle \dot{\phi}^2 \rangle = \frac{\omega}{2\pi}\int_0^{2\pi/\omega} \mathrm{d}t \left( \frac{\partial \phi}{\partial t} \right)^2\ , \\
    \langle \phi'^2 \rangle = \frac{\omega}{2\pi}\int_0^{2\pi/\omega} \mathrm{d}t \left( \frac{\partial \phi}{\partial r} \right)^2\ , \\
    \langle V \rangle = \frac{\omega}{2\pi}\int_0^{2\pi/\omega} \mathrm{d}t \ V(\phi)\ .
    \label{avg}
\end{split}
\end{equation}
Using Eqs. (\ref{deriv},\ref{avg}), the energy of the quasibreather can be defined as the time average
\begin{eqnarray}
    &&E_{QB} = \langle E \rangle
    \\
    &&= \frac{2\pi^{\frac{D}{2}}}{\Gamma\left(\frac{D}{2}\right)} \int_0^R \mathrm{d}r\ r^{D-1}\ \left[ \frac{1}{2} \langle \dot{\phi}^2 \rangle + \frac{1}{2}\langle \phi'^2 \rangle + \langle V \rangle \right]\ ,\nonumber
\end{eqnarray}
where $R$ is some cutoff coordinate that must be chosen outside the core region of the quasi-breather. Its precise  value is unimportant due to the small energy density of the quasibreather configuration at large values of $r$.
\clearpage

\end{document}